\documentclass[twocolumn,english,prx]{revtex4-1}
\usepackage{color}
\usepackage[utf8]{inputenc}
\usepackage{graphicx}
\usepackage[T1]{fontenc}
\usepackage{float}
\usepackage{natbib}
\usepackage{textcomp}
\usepackage{amsmath}
\usepackage{amssymb}
\usepackage{graphicx}
\usepackage{esint}
\usepackage{natbib}
\usepackage{xcolor}
\usepackage{multirow}
\usepackage{ulem}

\begin{document}
\title{Electron spin inversion in gated silicene nanoribbons}

\author{Bart\l{}omiej Rzeszotarski}

\affiliation{AGH University of Science and Technology, Faculty of Physics and
Applied Computer Science,\\
 al. Mickiewicza 30, 30-059 Kraków, Poland}

\author{Bart\l{}omiej Szafran}

\affiliation{AGH University of Science and Technology, Faculty of Physics and
Applied Computer Science,\\
 al. Mickiewicza 30, 30-059 Kraków, Poland}

\begin{abstract}
We study locally gated silicene nanoribbons as spin active devices. 
We find that the gated segments of zigzag nanoribbons can be used for  inversion of  spins. The strong intrinsic spin-orbit coupling
 for low Fermi energy in presence of an external vertical electric field provides a fast spin precession around the axis perpendicular to the silicene plane. 
The spin inversion length can be as small as 10 nm. On the other hand in the armchair nanoribbons the spin inversion occurs
via the Rashba effect which is weak and the spin inversion lengths are of the order of $\mu$m.

\end{abstract}
\maketitle
\section{Introduction}

Silicene \cite{chow} is buckled graphene-like structure with strong spin-orbit (SO) coupling \cite{Liu11,Liu,Ezawa} belonging to the 2D-Xenes group \cite{xene} that potentially can be used in spintronic devices \cite{Zutic04}. In silicene systems numerous quantum effects have been predicted, including the quantum spin Hall effect \cite{Liu11}, anomalous Hall effect \cite{Ezawa} and its valley-polarized variant \cite{Pan14}. Also,  an appearance of giant magnetoresistance is expected \cite{Xu12,Rachel14}. Furthermore, in presence of the external perpendicular electric field topological phase transitions in the edge states are predicted \cite{Ezawa2012a,romera,Tabert13}. 
Spin-filtering applications are possible for silicene \cite{Tsai13,wu15,miso15,shak15,nunez16}  with
the local exchange field introduced by the ferromagnetic proximity effects \cite{an,0yoko,1wang,2saxena,3derak,4wang}
or magnetic impurities \cite{derak16,rzang}.
In recent research, the silicene field effect transistor that operates at room-temperature has been demonstrated \cite{Tao15} with Al$_2$O$_3$ dielectric substrate that only weakly modifies the free-standing silicene band structure near the Dirac points \cite{al2o3} in contrast to Ag substrates \cite{Aufray10,Feng12,Vogt12}.

In the present paper we study electron spin inverter in silicene that exploits the SO interactions. 
The Rashba spin-orbit interaction due to the vertical electric field generates an in-plane effective magnetic field \cite{Meier2007} that in III-V two-dimensional electron gas
induces precession of the spin that is injected perpendicular to the plane of confinement. However, we find that for 
 zigzag silicene nanoribbon the precession in the Rashba effective field is blocked by the intrinsic SO interaction that generates strong internal magnetic field along the $z$ axis \cite{cum} that stabilizes the spin and stops its precession. The effects of the intrinsic SO coupling \cite{cum} are lifted for the armchair edge that introduces the intervalley scattering. The spin-precession
in armchair ribbons 
is observed but since the Rashba SO interaction in slicene is weak the spin inversion lengths are of the order of 1 $\mu$m which may not be attractive from the point of the practical application.

We show that spin precession occurs very fast under intrinsic SO coupling in zigzag silicene nanoribbons when the spin is injected within the ribbon plane perpendicular to the electron momentum. The difference in wave vectors can be easily chosen by Fermi energy in presence of the external electric field. The spin precession length can be tuned by the electric fields to the values lower than 10 nm. 



\section{Theory}

\subsection{Hamiltonian}
In calculations we use the $\pi$ band tight-binding Hamiltonian \cite{Liu} for the free standing silicene which takes the form 
\begin{align}
H_0=&-t\sum_{\langle k,l\rangle \alpha }  c_{k \alpha}^\dagger c_{l \alpha}+e  \sum_{k,\alpha}F_z \ell_k c^\dagger_{k,\alpha}c_{k,\alpha},  \nonumber \\ 
& -i\frac{2}{3}\lambda_{R}^{int.} \sum_{\langle \langle  k,l \rangle \rangle \alpha,\beta } \mu_{kl} c^\dagger _{k\alpha}\left(\vec{\sigma}\times\vec{d}_{kl} \right)^z_{\alpha\beta} c_{l\beta} \nonumber \\ 
&+i\frac{\lambda_{SO}}{3\sqrt{3}} \sum_{\langle \langle k,l\rangle \rangle \alpha, \beta } \nu_{kl} c^\dagger_{k\alpha} \sigma^{z}_{\alpha,\beta}c_{l\beta} \nonumber \\
& +i\lambda_{R}^{ext.} (F_z) \sum_{\langle k,l \rangle\alpha,\beta}  c_{k\alpha}^\dagger \left(\vec{\sigma}\times\vec{d}_{kl} \right)^z_{\alpha\beta}  c_{l\beta} \nonumber \\
& + \sum_{k,\alpha,\beta}c^\dagger_{k,\alpha}(\mathbf{M}\cdot\pmb{\sigma})_{\alpha\beta}c_{k,\beta}, 
\label{eq:h0}
\end{align}

\noindent where $c_{k \alpha}^\dagger$ ($c_{k \alpha}$) is the creation (annihilation) operator for an electron on site $k$ with spin $\alpha$. Summation over $\langle k,l\rangle$ and $\langle\langle k,l\rangle\rangle $ stands for the nearest and next nearest neighbor ions, respectively. 
(i) The first term of the Hamiltonian describes the hoppings between nearest atoms with $t=1.6$ eV \cite{Liu,Ezawa}.
(ii) The second term includes electrostatic potential due to electric field $F_z$ perpendicular to the system with $\ell_k=\pm \frac{0.46\text{\AA}}{2} $ 
with $+$ ($-$) sign for the ions of the A (B) sublattice.
(iii) The third term describes the intrinsic Rashba interaction with parameter $\lambda_{R}^{int.}=0.7$ meV \cite{Liu,Ezawa} due to the built-in electric field that emerges from the vertical shift of the $A$ and $B$ sublattices in silicene, where ${\bf d}_{kl}=\frac{{\bf r}_l-{\bf r_k}}{|{\bf r}_l-{\bf r_k}|}$ is the position of $k$-th ion and ${\bf r_k}=(x_k,y_k,z_k)$, with the lattice constant $a=3.86$ \AA. The $\mu_{kl}=+1$ ($-1$) for $\langle \langle k,l\rangle \rangle$ ions within sublattice A (B). 
(iv) The fourth term represents the effective SO coupling with $\lambda_{SO}=3.9$ meV in the Kane-Mele form \cite{km,km2} with $\nu_{kl}=+1$  ($-1$) for the counterclockwise (clockwise) next-nearest neighbor hopping.
(v) The fifth term describes the extrinsic Rashba effect which results from the external electric field perpendicular to the silicene plane or broken mirror symmetry by e.g. the substrate.
The parameter $\lambda_{R}^{ext.}$ varies linearly with the external field and for $F_z=17$ meV/\AA\; the $\lambda_{R}^{ext.}(F_z)=10$ $\mu$eV \cite{Ezawa}.
(vi) The last term introduces the local exchange field with magnetization described by exchange field $\mathbf{M}=(M_x,M_y,M_z)$ that may arise due to proximity of an insulating ferromagnetic substrate \cite{chow,an,0yoko,1wang,2saxena,3derak,4wang}.

To solve the scattering problem for the atomistic system described by Hamiltonian (\ref{eq:h0}) we use the wave function matching (WFM) technique as described in the appendix of Ref. \cite{bubel}. 
The transmission probability from the input lead to mode $m$ (output lead)
cab be written as
\begin{equation}
T^m = \sum_n \vert t^{mn} \vert ^2,
\end{equation}

\noindent where $t^{mn}$ is the probability amplitude for the transmission from the mode $n$ in the input lead to mode $m$ in the output lead. We distinguish spin for each mode $p$ by quantum expectation values of the Pauli matrices $\langle S_\bullet \rangle = \langle \psi_i^p \vert \sigma_\bullet \vert \psi_i^p \rangle$ through each atom $i$ inside lead. 
The positive (negative) $\langle S_\bullet \rangle$ values are labeled by $u$,$\uparrow$ ($d$,$\downarrow$). With this notation the spin-dependent conductance can be put in form 
\begin{equation}
G_{wv} = G_0 \sum_{m,n}  \vert t^{mn} \vert ^2 \delta_{w,\alpha(n)}\delta_{v,\beta(m)},
\end{equation}
\noindent where  $G_0 = e^2/h$ is the conductance quantum,  $w$ ($v$) is the expected input (output) orientation of the spin, while $\alpha$ and $\beta$ correspond to determined sings of $\langle S_\bullet \rangle$ sign for a given mode.
For example, for the incident spin polarized along the $z$ direction, the 
$\langle S_z \rangle = \langle \psi_i^p \vert \sigma_z \vert \psi_i^p \rangle$ is evaluated and 
the contribution to conductance that corresponds to the spin flip from $u$ to $d$ orientation is calculated as $G_{ud} = G_0 \sum_{m,n}  \vert t^{mn} \vert ^2 \delta_{+,\alpha(n)}\delta_{-,\beta(m)}$. 
All other spin-dependent conductance components can be calculated in the same way.
%

%

\begin{figure}[htbp]
\centering
\includegraphics[width=0.45\textwidth]{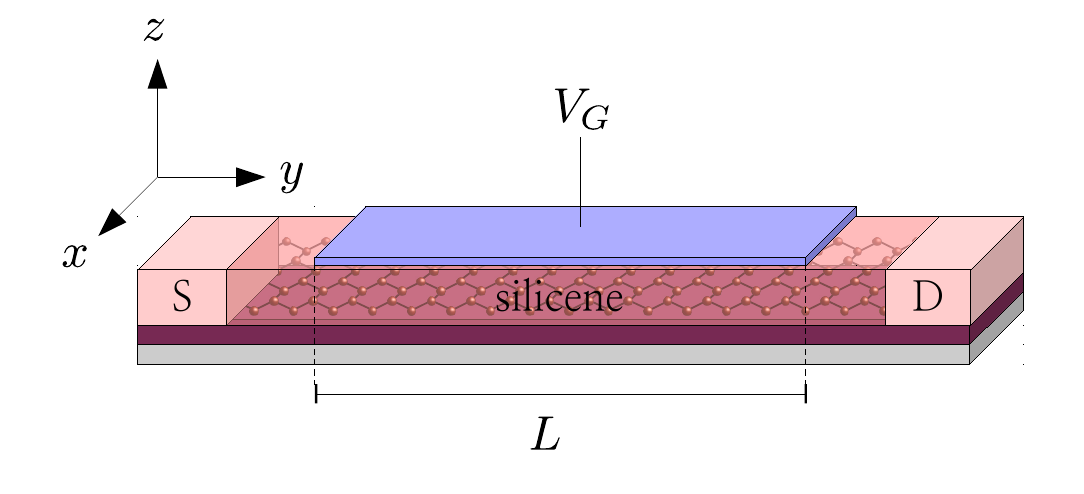}
\caption{ Sketch of the device. The Fermi level electrons propagate from the source (S) to the drain (D). The voltage $V_G$ applied to the top gate produces a perpendicular  electric field $F_z$  on length $L$ (see Eq. \ref{eq:h0}). The role 
of the source (S) and drain (D) is played by homogeneous semi-infinite silicene ribbons outside the gated area. }
\label{fig:0-sch}
\end{figure}

\begin{figure*}[htbp]
\centering
\begin{tabular}{c c}
\put(100,10){${\bf B}=(0,0,b)$ }
\put(250,10){${\bf B}=(b,0,0)$ }& \multirow{3}{*}{\includegraphics[clip,trim=0cm 1.2cm 0cm -1.2cm,scale=0.5]{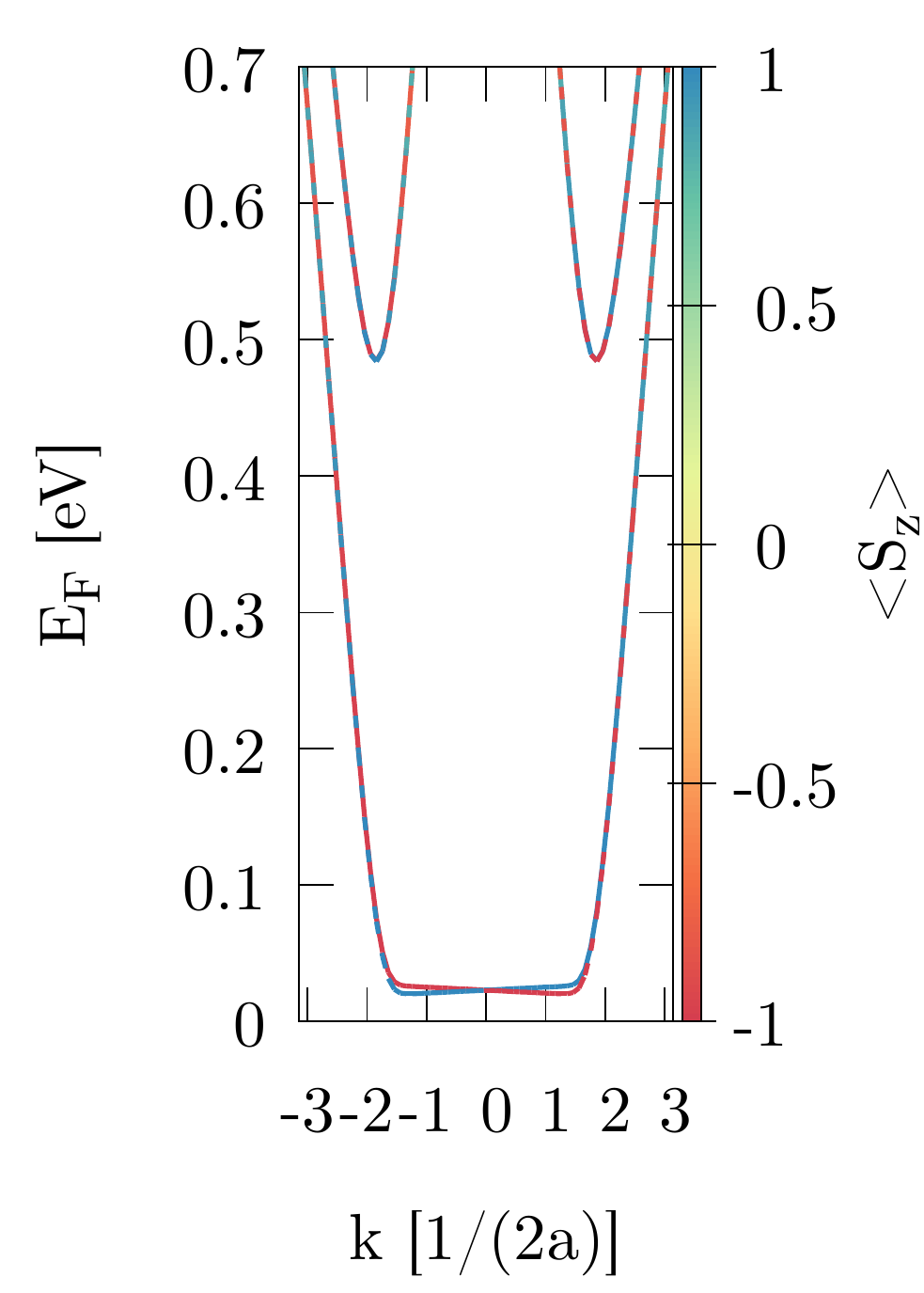}\put(-75,-10){k $\big[\frac{1}{2a}\big]$}\put(-95,160){(i)}} \\
\includegraphics[clip,trim=0cm 2.2cm 2.2cm 0cm,scale=0.5]{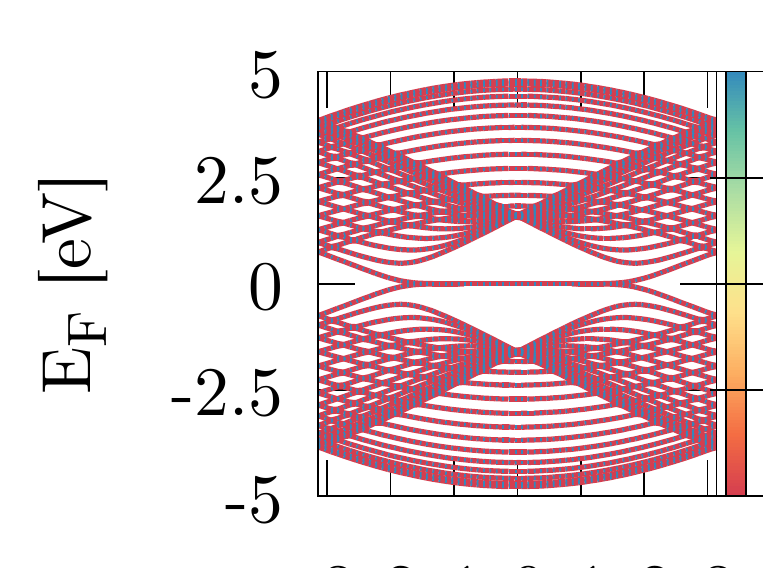}\put(-65,63){(a)}
\put(-65,80){$F_z = 0$ mV/\AA}
\put(5,80){$F_z = 100$ mV/\AA}
\includegraphics[clip,trim=2.8cm 2.2cm 2.2cm 0cm,scale=0.5]{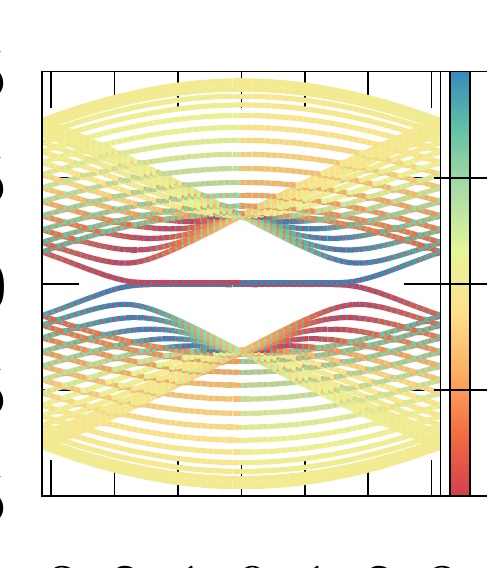}\put(-65,63){(b)}
\includegraphics[clip,trim=2.8cm 2.2cm 2.2cm 0cm,scale=0.5]{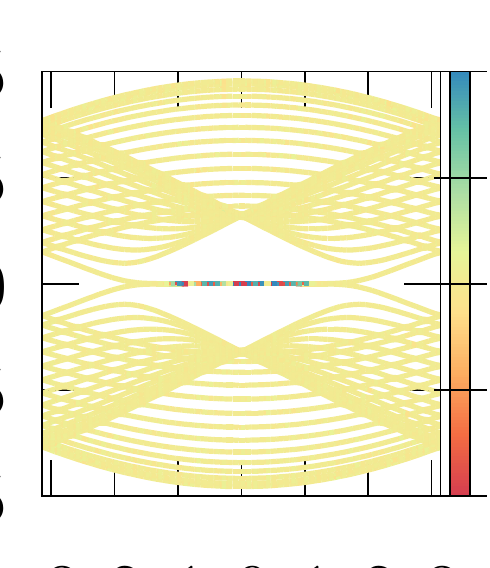}\put(-65,63){(c)}
\put(-65,80){$F_z = 0$ mV/\AA}
\put(5,80){$F_z = 100$ mV/\AA}
\includegraphics[clip,trim=2.8cm 2.2cm 0cm 0cm,scale=0.5]{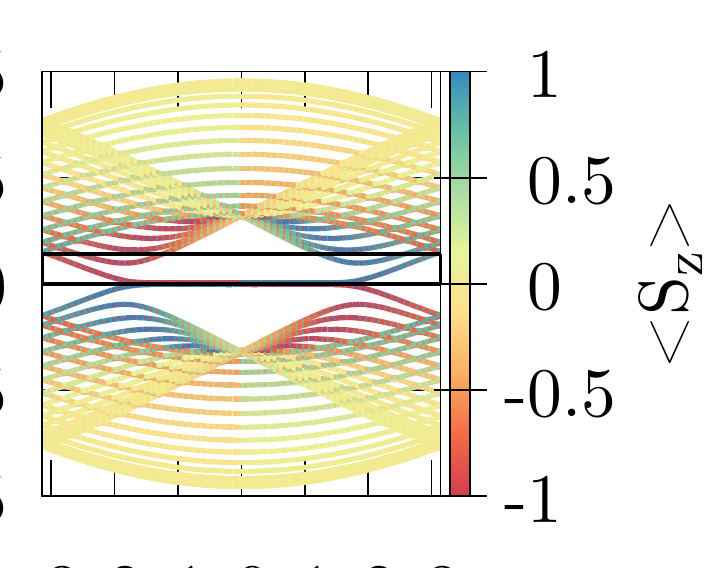}\put(-95,63){(d)}& \\
\includegraphics[clip,trim=0cm 1.4cm 2.2cm 0cm,scale=0.5]{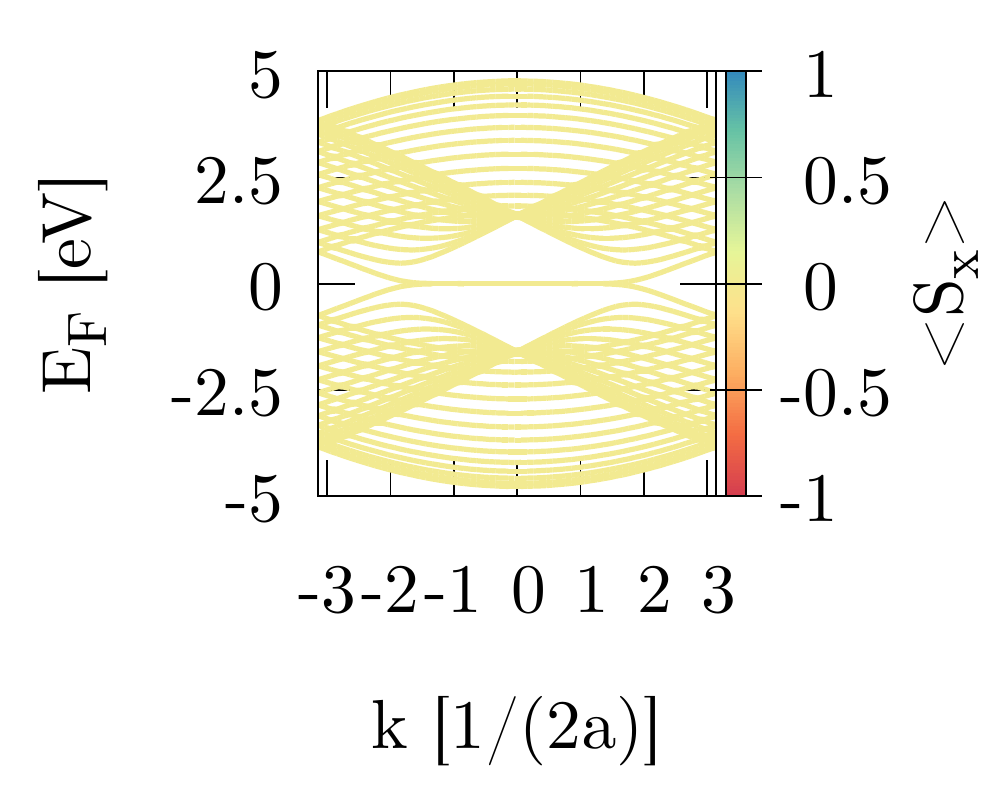}\put(-65,74){(e)}
\includegraphics[clip,trim=2.8cm 1.4cm 2.2cm 0cm,scale=0.5]{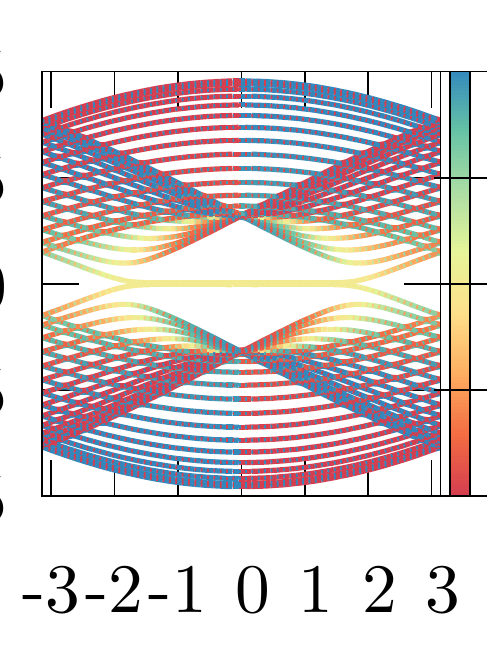}\put(-65,74){(f)}
\put(-50,-10){k $\big[\frac{1}{2a}\big]$}
\put(22,-10){k $\big[\frac{1}{2a}\big]$}
\put(-122,-10){k $\big[\frac{1}{2a}\big]$}
\put(95,-10){k $\big[\frac{1}{2a}\big]$}
\includegraphics[clip,trim=2.8cm 1.4cm 2.2cm 0cm,scale=0.5]{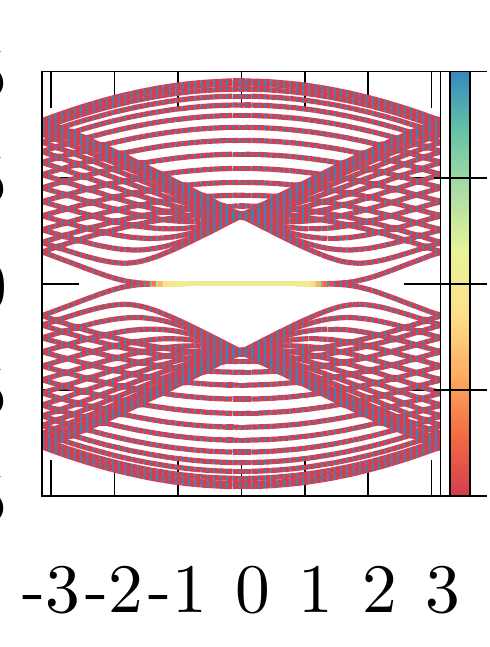}\put(-65,74){(g)}
\includegraphics[clip,trim=2.8cm 1.4cm 0cm 0cm,scale=0.5]{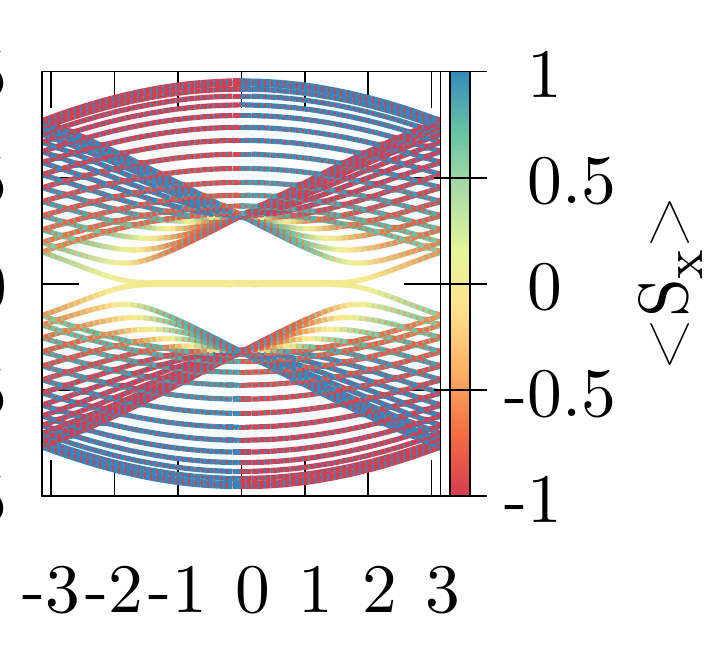}\put(-95,74){(h)}&  \\
\end{tabular}
\caption{ Band structure of a zigzag silicene nanoribbon with 28 atoms across its width.  Color in the upper a-d (lower e-h) row denote the spin $S_z$ ($S_x$) component in $\hbar/2$ units. An external magnetic field of $b=1\mu$T oriented in the $z$ and $x$ direction was applied in (a,b,e,f) and (c,d,g,h) respectively.
 Plots (a,e,c,g) represent band structures when no external electric field is applied while on the (b,f,d,h) the $F_z=100$ mV/\AA.   (i) Zoom of a band structure marked by rectangle in plot (d). }
\label{fig:1-bands}
\end{figure*}
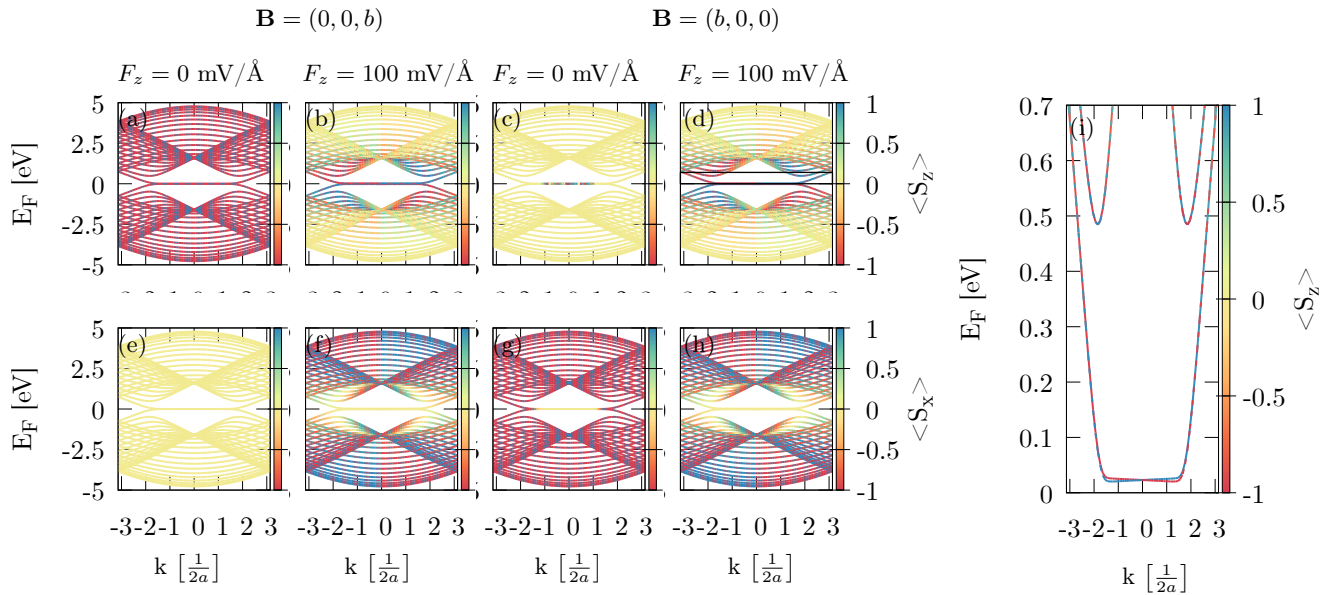

We consider the vertical orientation of the system in $z=0$ plane  where incident electron momentum is set in $y$ direction and the width of the ribbon is defined along the $x$ axis [see Fig. \ref{fig:0-sch}]. External electric field is applied perpendicular to system (along $z$ direction). 

\section{Results and Discussion}

\subsection{Perpendicular spin polarization}

\subsubsection{Zigzag lead}
In Fig. \ref{fig:1-bands} we show the dispersion relation and the average value of the spin for a zigzag silicene ribbon covered by an {\it infinite} top gate [Fig. \ref{fig:0-sch}] that can produce a homogeneous vertical electric field.  
The ribbon is 28 atoms wide, which corresponds to a width of $\simeq 5$ nm. 
Note, that such narrow or even narrower -- down to 0.8 nm -- silicene nanoribbons 
have been grown experimentally \cite{hiraoka}.

In order to probe the spin properties of the system we introduced a very small external magnetic field equal to 1 $\mu$T oriented in the $z$ [Fig. 2(a,b,e,f)] or $x$ [Fig. 2(c,d,g,h)]  direction. 
In Fig. \ref{fig:1-bands} all the bands are nearly two-fold degenerate with respect to the spin. The lifting of the spin degeneracy
 can only
be observed on enlarged fragments which are discussed in detail below in the text.  

For ${\bf B}=(0,0,b)$ with no external electric field the electron spin is polarized in the $\pm z$ direction for any Fermi energy [Fig. \ref{fig:1-bands}(a,e)]. 
When the external electric field of $F_z = 100$ mV/\AA{} is switched on,  the external Rashba interaction introduces 
an effective magnetic field  $\mathbf{B}_{\lambda_R^{ext.}}= \xi ( \mathbf{p \times E} )= (\xi p_yF_z,0,0) $ \cite{Meier2007} ($\xi$ is a constant) 
that tends to set the spins parallel or antiparallel to the $x$ axis. 
On the other hand the intrinsic spin-orbit coupling of the Kane-Mele form in the absence of the intervalley scattering, tends to polarize the spins in the direction perpendicular to the silicene plane  \cite{cum}.
Figs. 2(b) and 2(f) illustrate the competition between the intrinsic spin-orbit coupling and the external Rashba interaction. The latter prevails for high Fermi energy
for which the spins are polarized within silicene plane in the $\pm x$ direction. 
For lower Fermi energy the spin-diagonal  intrinsic SO coupling keeps the spin polarized in  $\pm z$ direction \cite{cum}.

\begin{figure}[htbp]
\centering
\includegraphics[clip,trim=0cm 0cm 0cm 0cm,scale=0.4]{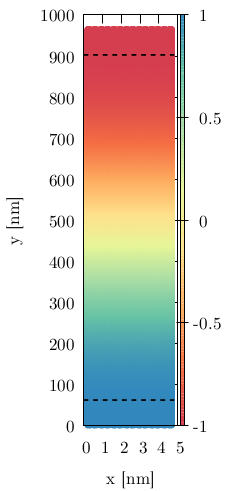}
\includegraphics[clip,trim=0cm 0cm 0cm 0cm,scale=0.4]{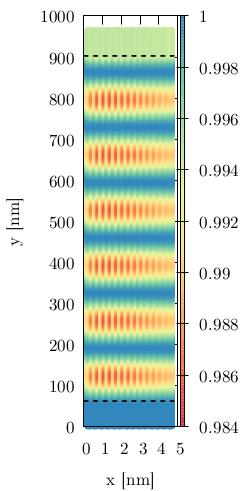}
\put(-200,200){(a)}
\put(-100,200){(b)}
\caption{ Spin $S_z$ component maps for zigzag nanoribbon 28 atoms wide for  the intrinsic SO coupling constant $\lambda_{SO}=0$ (a)  and  $\lambda_{SO}=3.9$ meV (b). In between dashed horizontal lines the external electric field $F_z=1$ V/\AA{} is applied. The first propagating mode with initial spin up has been chosen at $E_F=258$ meV. }
\label{fig:2-maps}
\end{figure}

Now, let us  consider a device, i.e. a system with a {\it finite} top gate [see Fig. 1] and the spin injected polarized in the $z$ direction.
The scattering problem for the zigzag nanoribbon has been solved for finite length of the external electric field $F_z=1$ V/\AA{} at $E_F=0.258$ eV in two cases:   
{\it (i)} For $\lambda_{SO}=0$ the splitting between the spin-polarized subbands is $\Delta k_1=0.00288 \frac{1}{2a}$.
The Rashba SO interaction induces the spin precession with respect to the $x$ axis. The rotation of the electron spin along the $x$ axis upon transition along the length of $L_1$ can be evaluated as \cite{Datta90}
\begin{equation}
\Delta \varphi = \Delta k L_1.
\label{eq:dat}
\end{equation}
For $L_1(\pi)=842$ nm the spin rotates from $d$ to $u$ orientation [Fig. \ref{fig:2-maps}(a)].
{\it (ii) } For $\lambda_{SO}=3.9$ meV the intrinsic spin-orbit coupling keeps the electron spin polarized along the $z$
direction with only weak oscillations due to $\Delta k_2=0.036 \frac{1}{2a}$ and the $L_2(\pi)=67.4$ nm [Fig. \ref{fig:2-maps}(b)].

We conclude that the realization of a perpendicular spin  inverter in a zigzag  nanoribbon at low Fermi energy is excluded by the strong intrinsic SO which keeps the  spin polarized along the  $z$ axis.
\subsubsection{Armchair lead}
\begin{figure}[htbp]
\centering
\includegraphics[clip,trim=0cm 2.5cm 2.2cm 0.5cm,scale=0.5]{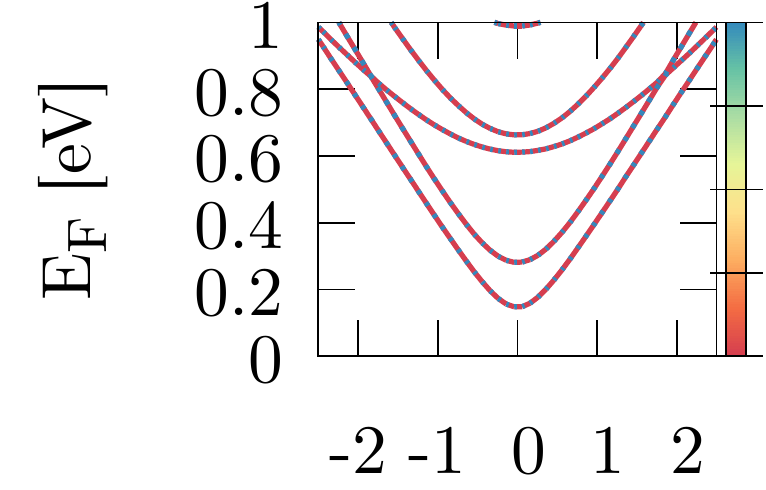}
\put(-65,78){$F_z = 0$ V/\AA}
\put(10,78){$F_z = 1$ V/\AA}
\put(-95,75){(a)}
\put(90,75){(b)}
\includegraphics[clip,trim=3cm 2.5cm 0cm 0.5cm,scale=0.5]{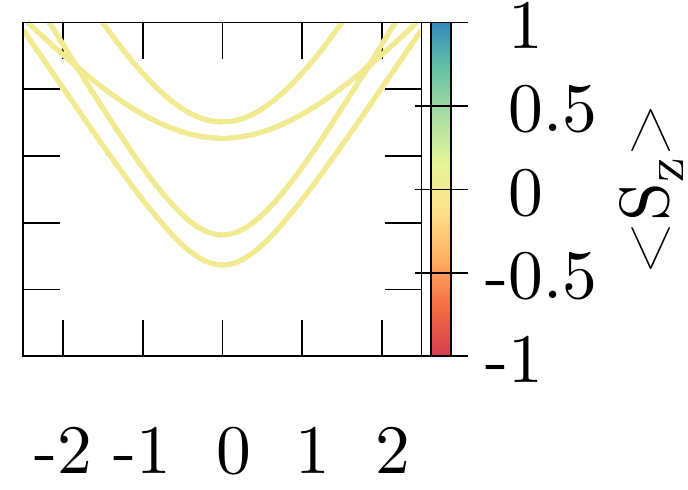}
\includegraphics[clip,trim=0cm 2cm 2.2cm 0.5cm,scale=0.5]{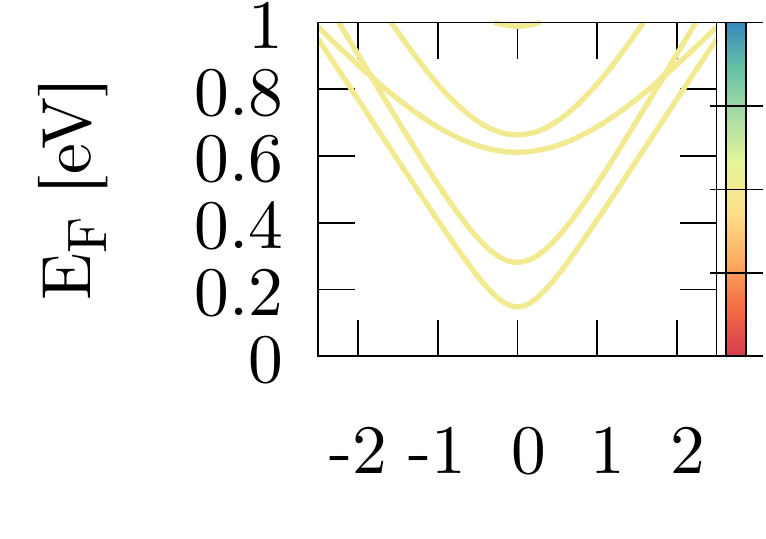}
\put(-50,-10){k $\big[\frac{1}{6a_{Si}}\big]$}
\put(22,-10){k $\big[\frac{1}{6a_{Si}}\big]$}
\put(-95,75){(c)}
\put(90,75){(d)}
\includegraphics[clip,trim=3cm 2cm 0cm 0.5cm,scale=0.5]{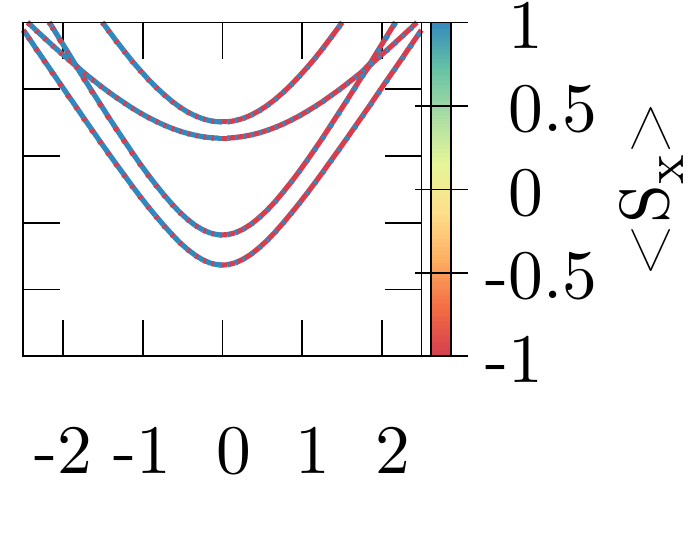}
\caption{ Band structure of an armchair silicene nanoribbon with 19 atoms width for electron with spin polarization in $z$ direction. Colors in upper (lower) row represent spin $S_z$ ($S_x$) component in $\hbar/2$ units. The left panels correspond to external electric field $F_z = 0$ V/\AA \; and the right panels to  $F_z = 1$ V/\AA\;. }
\label{fig:3-bands_z}
\end{figure}

\begin{figure}[htbp]
\centering
\includegraphics[width=0.5\textwidth]{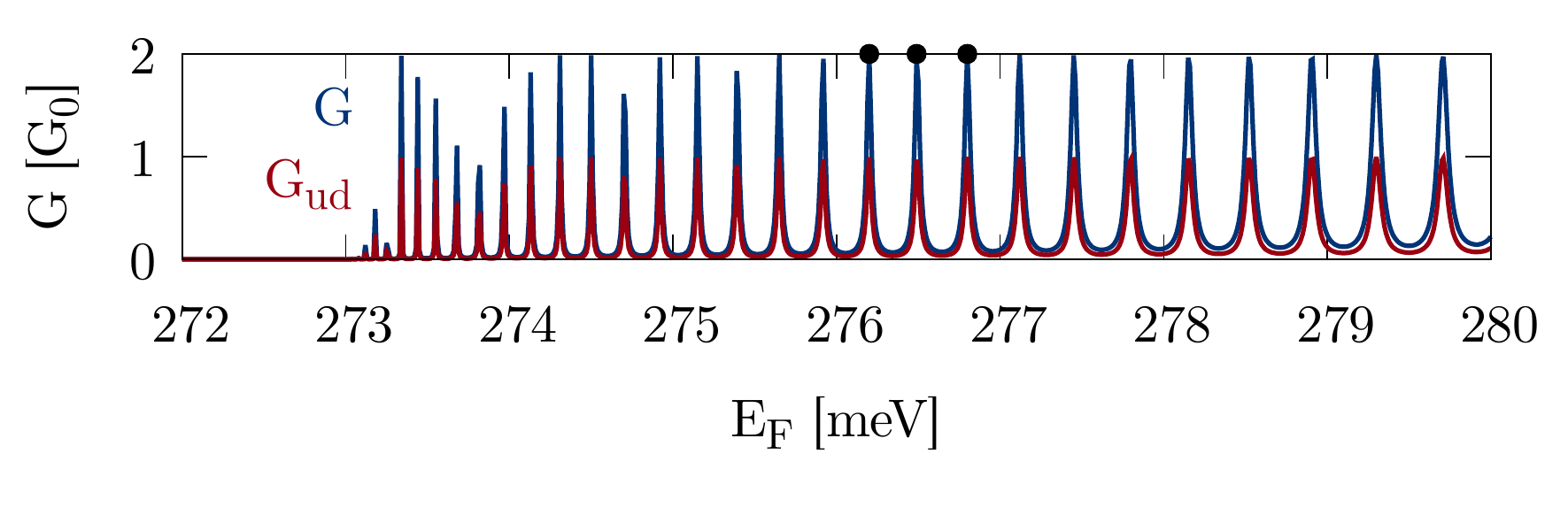}
\caption{ Total conductance $G$ (blue) and the spin-flipping $G_{ud}$ contribution (red) for an armchair silicene nanoribbon with the length of the gated area $L=842$ nm and the external electric field $F_z=1$ V/\AA\. 
The basis of spin states perpendicular to the ribbon is considered here.  The black dots marked the peaks described in Tab. \ref{tab:wl}.  }
\label{fig:3-trans}
\end{figure}

The effective magnetic field due to the intrinsic spin-orbit interaction that prevents the spin precession 
along the Rashba effective field is only present provided that the transport modes have a definite valley  \cite{cum,km}. The armchair edge of the ribbon 
introduces maximal valley mixing and removes the intrinsic spin-orbit effective magnetic field. 
In order to eliminate the effects of the intrinsic spin-orbit interaction we considered an armchair semiconducting ribbon 19 atoms wide (for metallic version precession occurs in the same manner). In the armchair nanoribbon with no electric field the initial spin $z$ is conserved [Fig. \ref{fig:3-bands_z}(a,c)] but when a high electric field  $F_z=1$ V/\AA{} is applied, the available states correspond to spins polarized along $x$ axis [Fig. \ref{fig:3-bands_z}(b,d)] due to the Rashba effective magnetic field. 
\indent The conductance for $L=842$ nm and $F_z=1$ V/\AA\ was calculated and presented in Fig. \ref{fig:3-trans}. 
Total conductance (blue line) and spin-flipping conductance (red line) oscillates 
with peaks for integer number of wavelength halves within the gated length $L$ [see Tab. \ref{tab:wl}]. For $F_z\neq 0$ a step potential appears in silicene sublattices
that  allows transmission  for resonant modes only. 

\begin{table}[htbp]
\caption{Fermi wavelengths $\lambda_m$ in resonances [see black dots in Fig. \ref{fig:3-trans}]. 
The results are obtained from the band structure for armchair ribbon (19 atoms width) with vertical electric field $F_z=1$ V/\AA. 
$m_1$ and $m_2$ stand for the two modes in the first conductive subband at the Fermi level.}
\centering
\begin{tabular}{|c|c|c|c|c|c|}
\hline
$E_F$ [meV] &  & $k$ [1/6$a_{Si}$] & $\lambda_m=\frac{2\pi}{k}$ [nm] & $N=\frac{L}{\lambda_m}$ & $\overline{N}$  \\ \hline
\multirow{2}{*}{276.8} &$m_1$&0.1172&71.71&11.75&\multirow{2}{*}{12}\\ \cline{2-5}
                       &$m_2$&0.1221&68.78&12.25&\\ \hline
\multirow{2}{*}{276.49}&$m_1$&0.1121&74.93&11.25&\multirow{2}{*}{11.5}\\ \cline{2-5}
                       &$m_2$&0.1171&71.74&11.75&\\ \hline
\multirow{2}{*}{276.2} &$m_1$&0.1072&78.37&10.75&\multirow{2}{*}{11}\\ \cline{2-5}
                       &$m_2$&0.1122&74.89&11.25&\\ \hline
\end{tabular}
\label{tab:wl}
\end{table}

Figure \ref{fig:3-trans} shows that for the armchair ribbon and the chosen length of the gated area the resonant electron transfer is accompanied by the spin-flip.
In a wide range around $E_F=276.49$ meV the spin-subbands splitting remains almost the same $\Delta k \approx 0.005 \frac{1}{6a_{Si}}$ (where $a_{Si}=\frac{a}{\sqrt{3}}$ is the in-plane distance between nearest-neighbors Si atoms, thus $L_a(\pi) \approx 842$ nm) and provides a perfect spin inverter. 
We can see that the spin inversion length is very large even for an extreme value 1V/\AA\; of the electric field applied here \cite{drumm,Ni2011}, which is not promising in the context of practical applications.

\subsection{In-plane spin polarization}

\begin{figure}[htbp]
\centering
\includegraphics[clip,trim=0cm 1cm 0cm 0cm,width=0.5\textwidth]{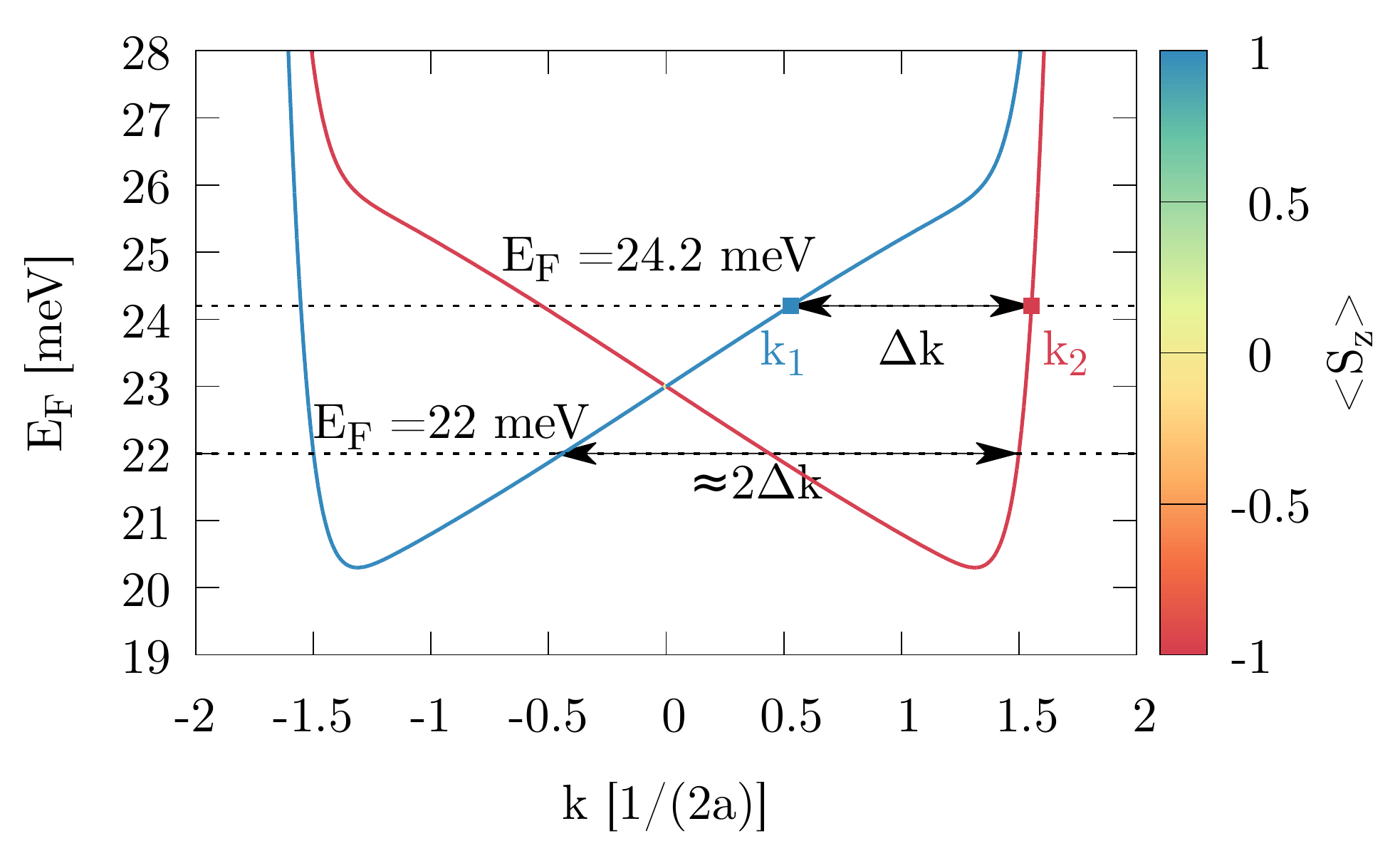}
\put(-145,-10){k $\big[\frac{1}{2a}\big]$}
\caption{ The lowest conduction subbands for zigzag nanoribbon and $F_z=100$ mV/\AA\;. The color scale denotes the spin $S_z$ component in $\hbar/2$ units. The difference between two opposite-spin right-going subbands at Fermi energy $E_F = 24.2$ meV is marked as $\Delta k = 1.02\; \frac{1}{2a}$.}
\label{fig:4-agressive}
\end{figure}

Figure \ref{fig:1-bands}(c,g) shows that for the electrons fed from the silicene lead in which the external electric field is absent 
one can polarize the spins in the $x$ direction with an infinitesimal magnetic field.  For  $F_z=0$, above energy $E_F=10$ meV, the electron spin is oriented parallel or antiparallel to the $x$ axis
by the field of 1 $\mu$T.  Moreover,  Fig. 2(d) indicates that in presence of nonzero $F_z$ for low Fermi energy the spins of the transport modes are polarized along the $z$ axis.
One can use this fact in order to arrange for a device which inverts the in-plane polarized incident spins for the electrons that enter a gated region.
 Furthermore if we apply external electric field $F_z=100$ mV/\AA, the first subbands [see Fig. \ref{fig:1-bands}(i)] with specified spin $z$ states splits in a very wide $\Delta k$ spectrum [zoom in Fig. \ref{fig:4-agressive}] between the two propagating modes.


\begin{figure}[htbp]
\centering
\includegraphics[width=0.4\textwidth]{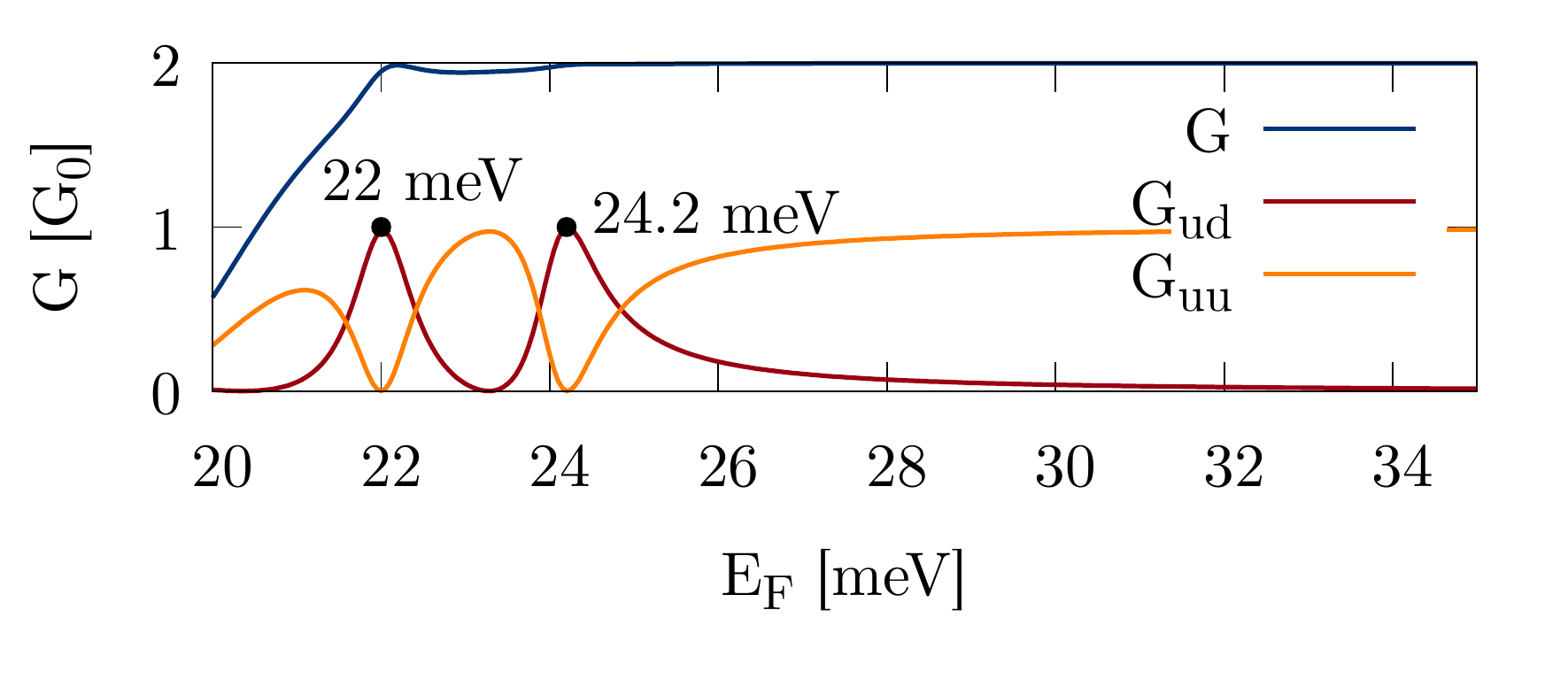}
\caption{ Total conductance $G$ (blue), the spin flipping  $G_{ud}$ (red) and spin conserving $G_{uu}$ (orange) contributions. Here, $G_{ud}$ denotes the flip from $S_x=1$ to $S_x=-1$ in $\hbar/2$ units. The width of the zigzag ribbon was set to 28 atoms, the  length is $L=5.5$ nm and $F_z=100$ mV/\AA. The two peaks marked by black dots correspond to dashed lines in band structure in Fig. \ref{fig:4-agressive}.}
\label{fig:4-trans}
\end{figure}

Figure \ref{fig:4-trans} shows the conductance for the zigzag nanoribbon (28 atoms width) with external electric field $F_z=100$ mV/\AA\ and $L=5.5$ nm.
We show the total conductance $G$ and its the spin-resolved contributions to conductance
$G=G_{ud}+G_{uu}+G_{du}+G_{dd}$.
$G_{ud}$ stands for the flip from $1$ to $-1$ in $S_x [\hbar/2]$ component, and $G_{uu}$ for the transport with the spin kept parallel
to the $x$ axis. Since the time-reversal symmetry is conserved one has $G_{du}=G_{ud}$ and  $G_{dd}=G_{uu}$. 

We find two peaks of $G_{ud}\simeq G_0 $ for $E_F=22$ meV and 24.2 meV. For the first peak the difference of the Fermi wave vectors is roughly
twice larger than in the other [Fig. 6].
For higher $E_F$ the $\Delta k$ values drastically drop and the precession is too slow to invert spin on that short $L$ path and $G_{ud}$ in Fig. 7 drops to zero.

\begin{figure}[htbp]
\centering
\includegraphics[clip,trim=0cm 2cm 0cm 0cm,scale=0.5]{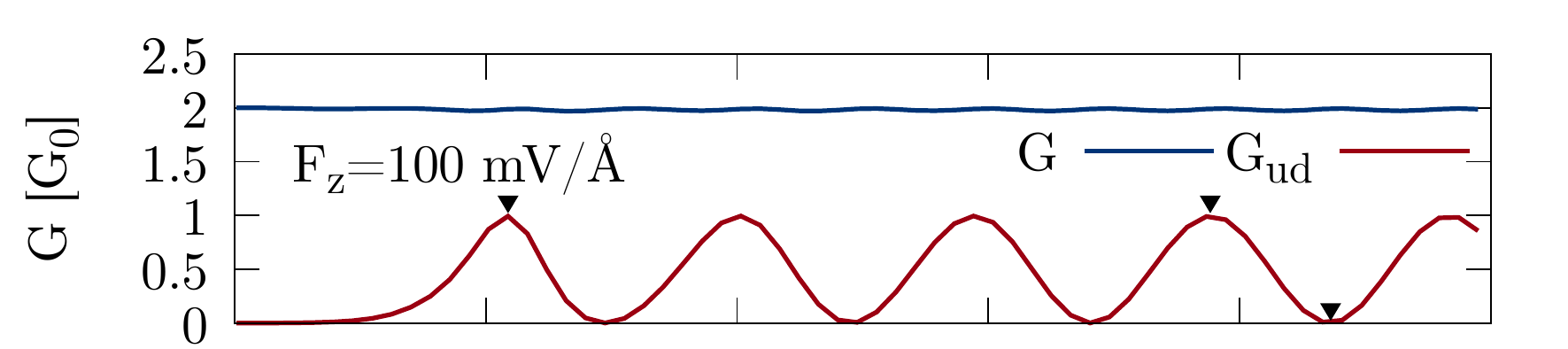}\put(-260,50){(a)}

\includegraphics[clip,trim=0cm 2cm 0cm 0cm,scale=0.5]{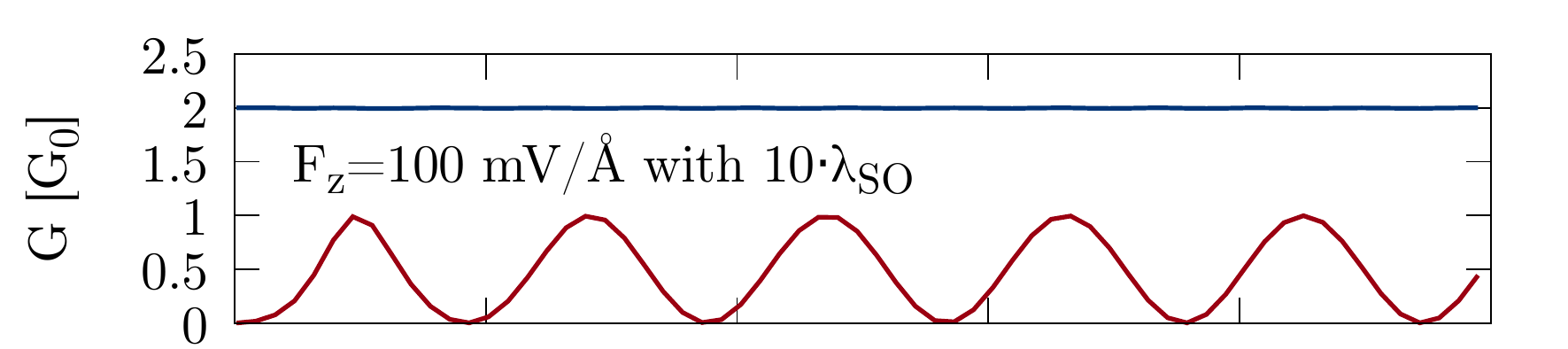}\put(-260,50){(b)}

\includegraphics[clip,trim=0cm 2cm 0cm 0cm,scale=0.5]{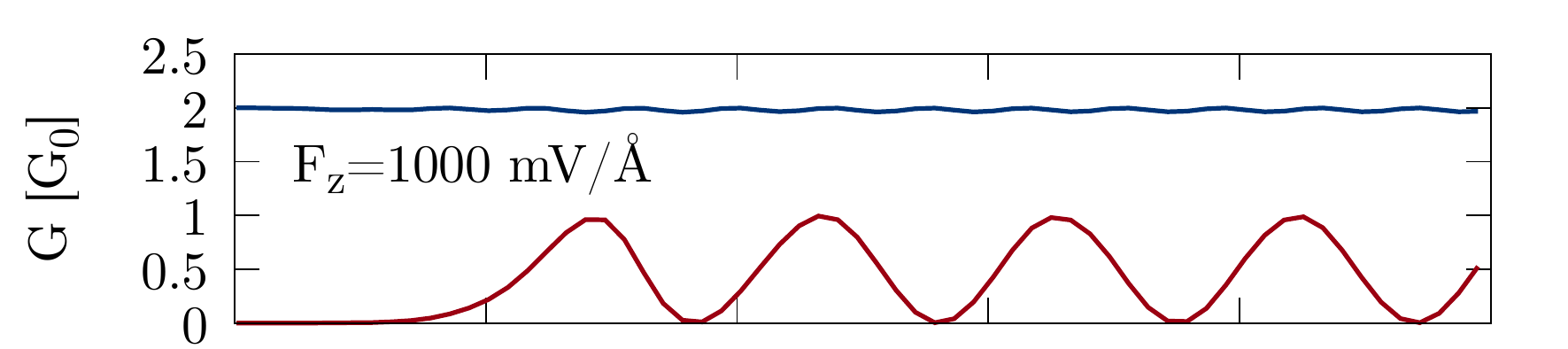}\put(-260,50){(c)}

\includegraphics[clip,trim=0cm 0cm 0cm 0cm,scale=0.5]{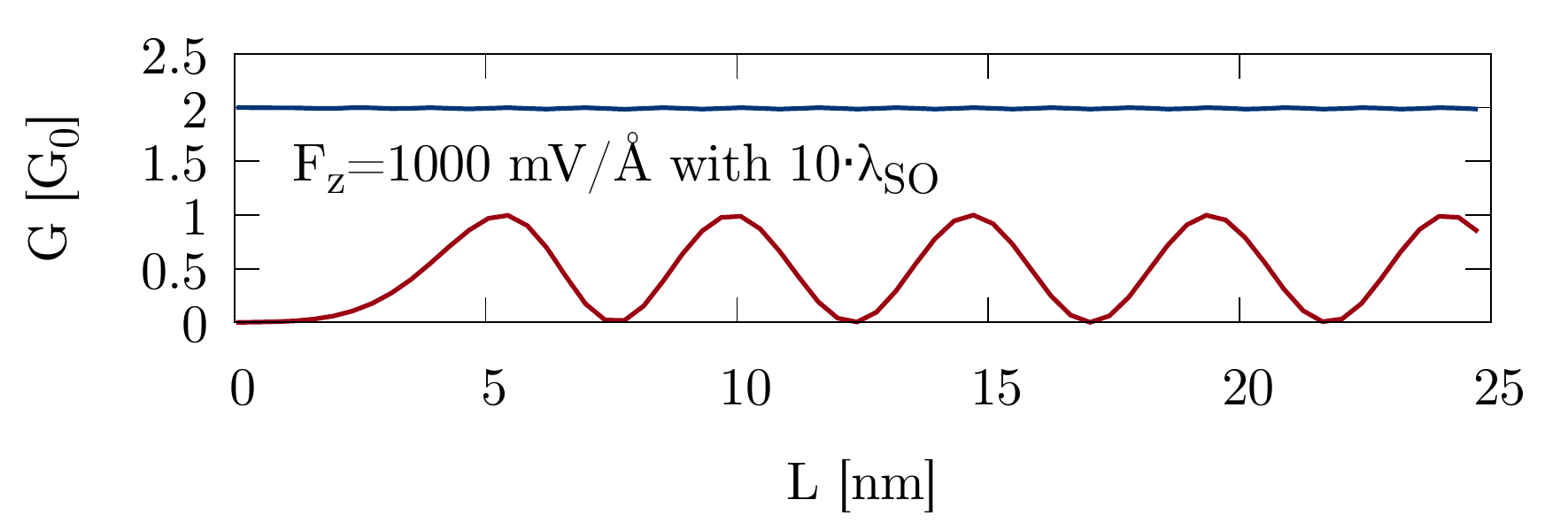}\put(-260,80){(d)}
\caption{ Total conductance $G$ (blue) and conductance with spin flip $G_{ud}$ (in $x$ direction) with variable length $L$ on which external electric field is applied. For (a,b) the $F_z=100$ mV/\AA\ and (c,d) $F_z=1000$ mV/\AA. In two cases (b,d) the $\lambda_{SO}$ was increased 10 times to measure the contribution of the intrinsic spin-orbit coupling to the offset arising. For black triangles in (a) - the first mark from the left corresponds to system described in text, the middle and the last one to Fig. \ref{fig:5-maps}(a,b), respectively. The applied Fermi energies: $E_{F}=24.2$  meV (a), $E_{F}=38.386$ meV (b), $E_{F}=231.653$ meV (c), and $E_{F}=252.989$ meV (d). }
\label{fig:5-offset}
\end{figure}

\begin{figure*}[htbp]
\centering
\includegraphics[clip,trim=3cm 0cm 5cm 2cm,scale=0.35]{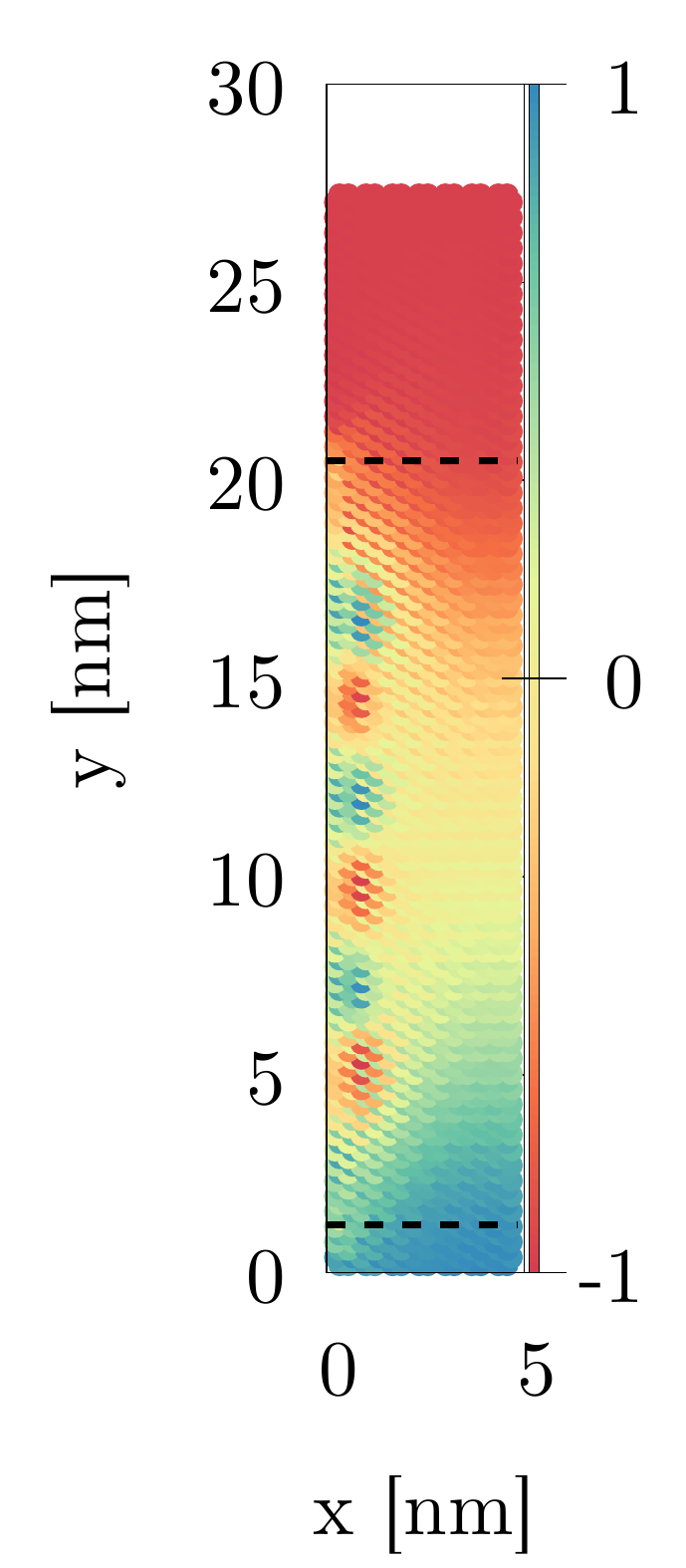}\put(-33,160){(a)}
\includegraphics[clip,trim=5cm 0cm 5cm 2cm,scale=0.35]{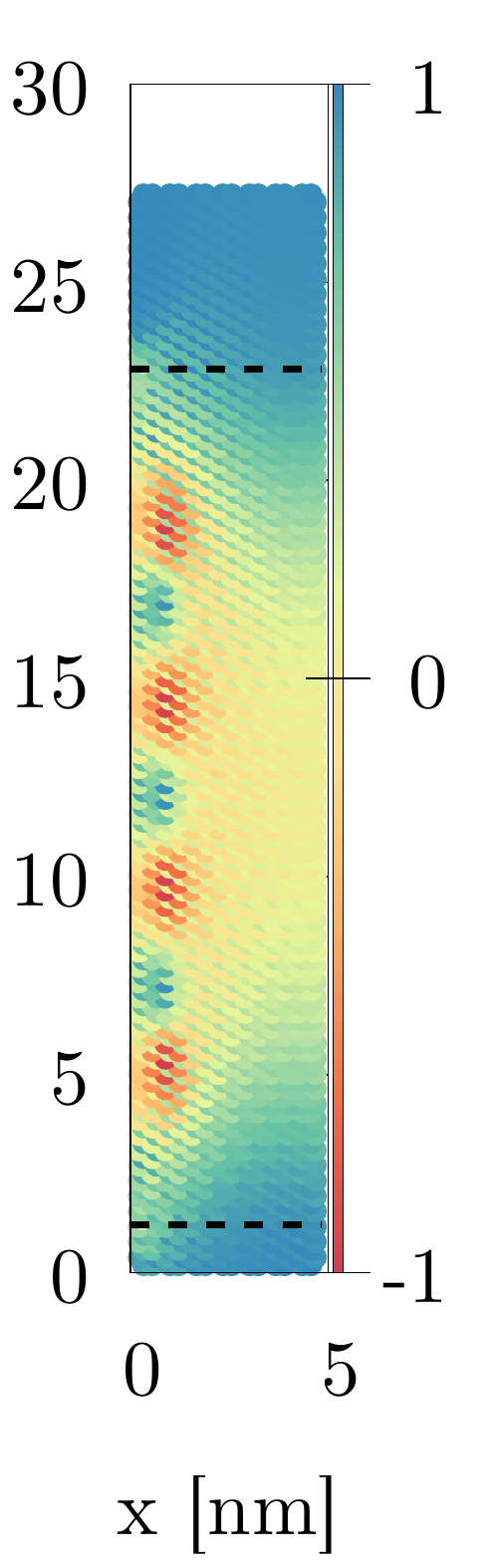}\put(-33,160){(b)}
\includegraphics[clip,trim=3cm 0cm 0cm 2cm,scale=0.35]{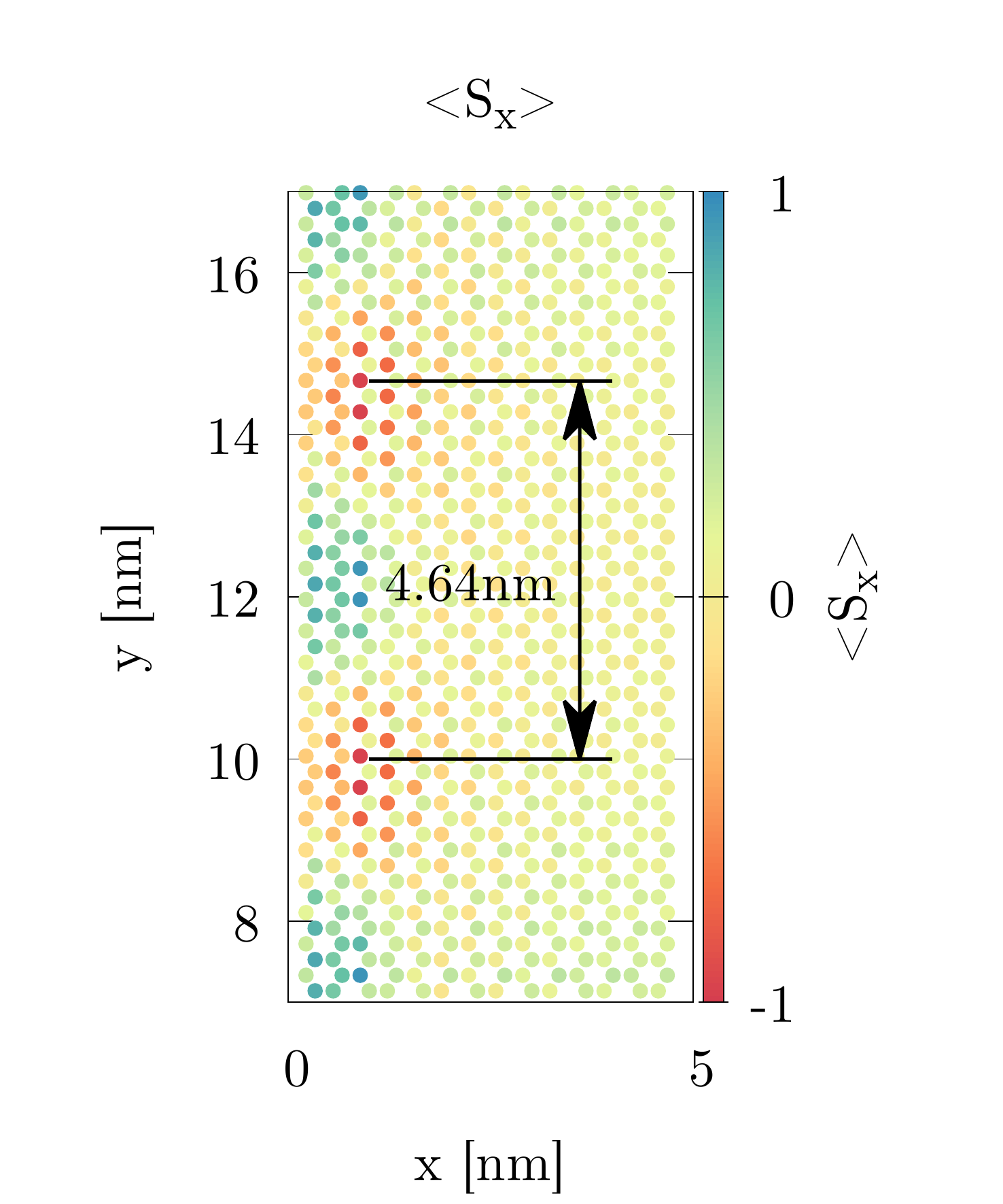}\put(-80,160){(c)}
\includegraphics[clip,trim=5cm 0cm 4cm 2cm,scale=0.35]{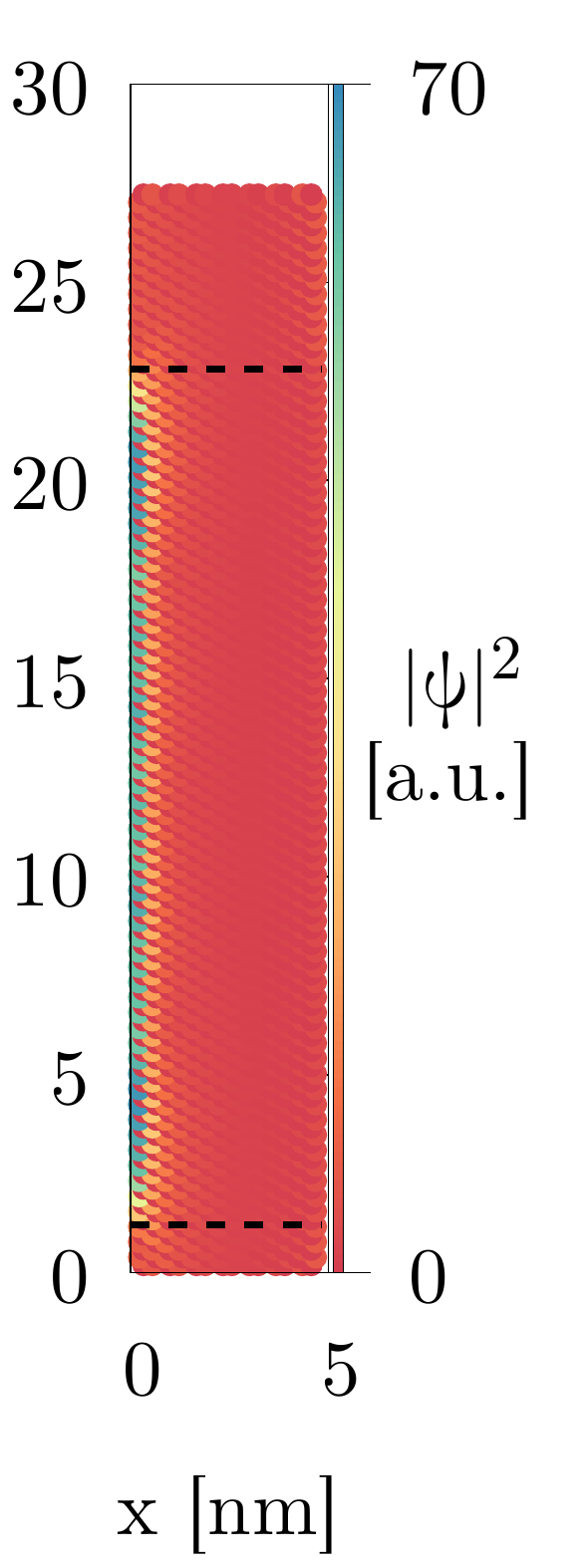}\put(-43,160){(d)}
\includegraphics[clip,trim=5cm 0cm 4cm 2cm,scale=0.35]{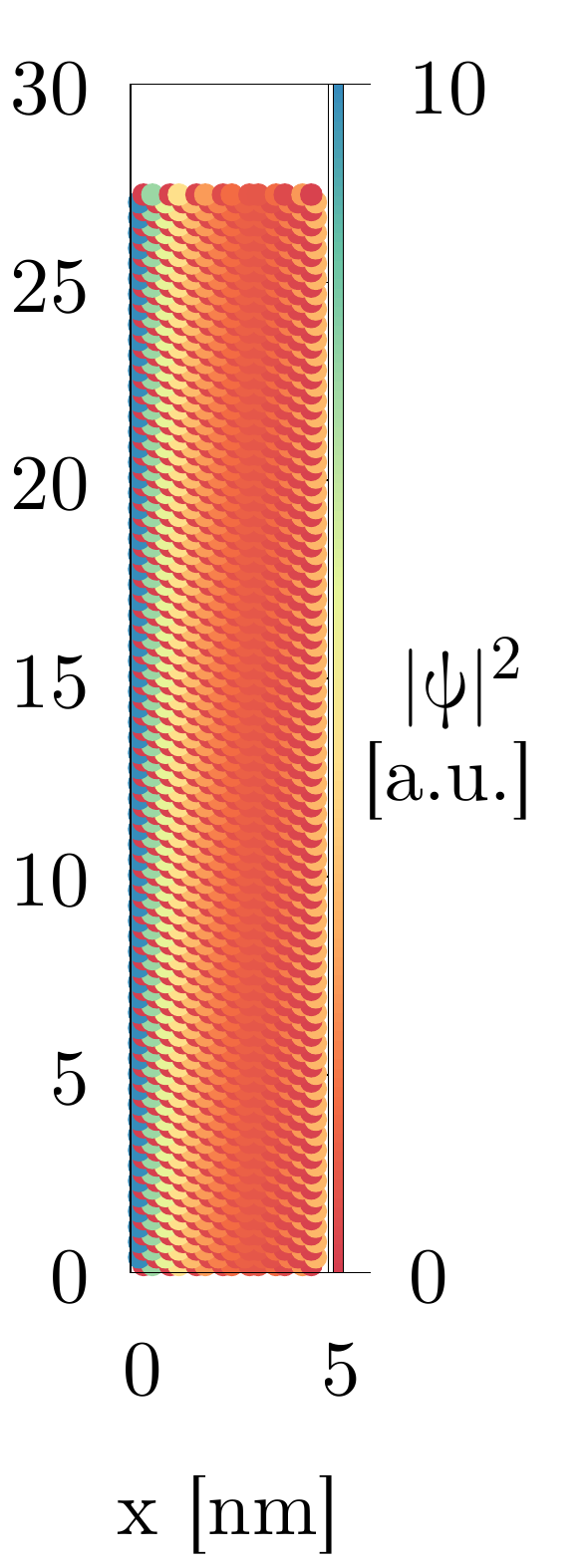}\put(-42,160){(e)}\put(-40,60){\rotatebox{90}{lead mode $m_1$}}
\includegraphics[clip,trim=5cm 0cm 4cm 2cm,scale=0.35]{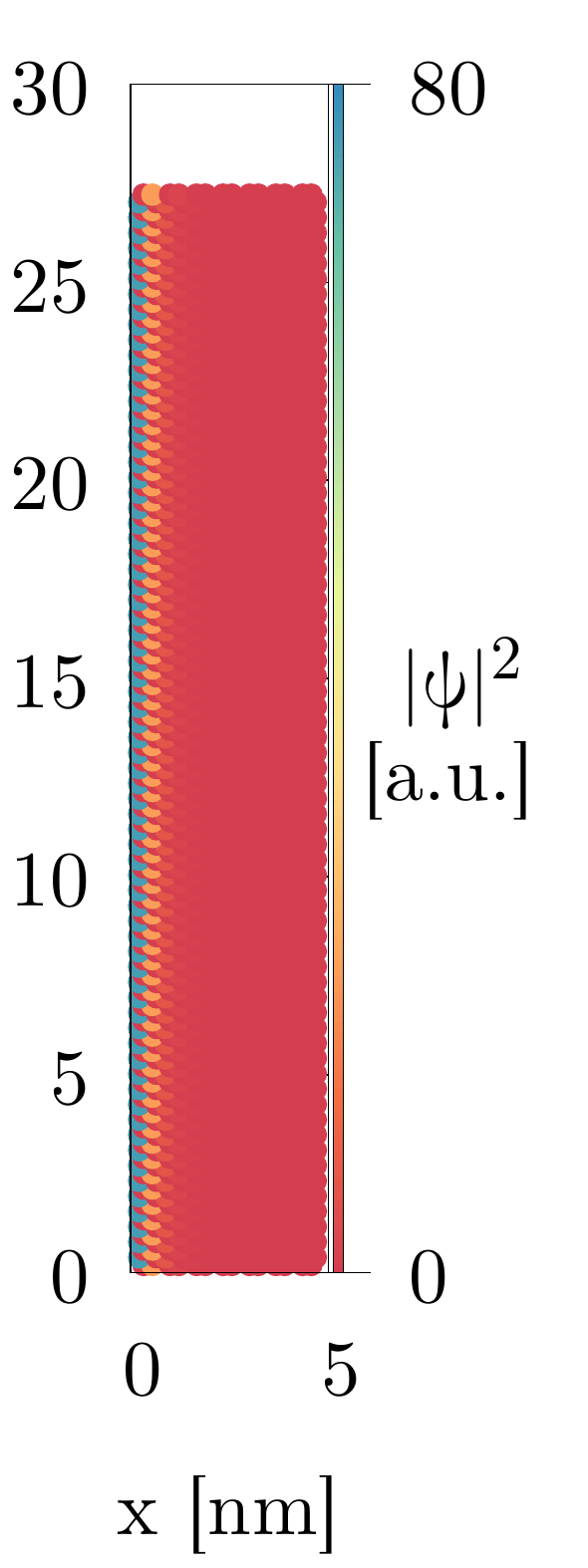}\put(-42,160){(f)}\put(-40,60){\rotatebox{90}{lead mode $m_2$}}
%
\caption{ $S_x$ spin projection maps for $L$ (a) 19.3nm and (b) 21.6nm [see the length marked by triangles in Fig. \ref{fig:5-offset}(a)]
for the solution of the scattering problem with spin-up electron incident from the side of negative $y$.
 (c) Zoom of the area where one period of the spin rotation is visualized. (d) Scattering density for system showed in plot (b). The horizontal dashed lines limit the area of the external electric field $F_z=100$ mV/\AA. The electron density of the input lead with $F_z=100$ mV/\AA{} as above  for two incoming modes $m_1$ (e) and $m_2$ (f).  }
\label{fig:5-maps}
\end{figure*}

For the $E_F=24.2$ meV the spacing of the right-going wave vectors for opposite spin states is  $\Delta k = 1.02\; \frac{1}{2a}$
 which according to Eq. (\ref{eq:dat}) provides the spin precession length $L_x(\pi) = 6a \approx 2.32$ nm. 
In Fig. \ref{fig:5-offset} we plotted the spin-flipping conductance as a function of the length of the gated region for varied values of the $F_z$ and $\lambda_{SO}$.
For each subplot in Fig. \ref{fig:5-offset} the Fermi energy was tuned to maintain the same spin precession rate: $\Delta k = 1.02\; \frac{1}{2a}$, $L_x(\pi)\approx 2.32$ nm.
We find that the subsequent peaks of the spin-flipping conductance $G_{ud}$ are  separated by  $2 L_x(\pi) = 4.64$ nm,
corresponding to an additional full spin rotation from one peak to the other.
The offset between the nominal $L_x(\pi)$ and the actual $L$ value for which the first peak of $G_{ud}$ occurs is due to the  finite size of the top gate.
Note, that the band structure of Fig. 6 is calculated for an infinite $L$. We find that this offset is dependent of the Fermi energy (the higher $E_F$, the lower wave sensitivity to the step of $F_z$ potential) and on intrinsic SO interaction strength. Increasing ten times the intrinsic SO factor $\lambda_{SO}$  we significantly shorten the offset lengths.  At the start and at the end of $F_z$ area the spin precession is unsettled which extends the actual spin inversion length [Fig. \ref{fig:5-maps}(a,b)]. 
In the scattering spin density the local extrema of $S_x$ are indeed spaced by $2\cdot L_x(\pi) = 4.64$ nm [Fig. \ref{fig:5-maps}(c)]. 
The fact that the scattering density mostly occupies the left edge of the nanoribbon [Fig. \ref{fig:5-maps}(d)] is consistent with the results obtained for infinite $L$ [Fig. \ref{fig:5-maps}(e,f)].

\begin{figure*}[htbp]
\centering
\includegraphics[clip,trim=-0.5cm 0.7cm 0cm 0cm,width=0.65\textwidth]{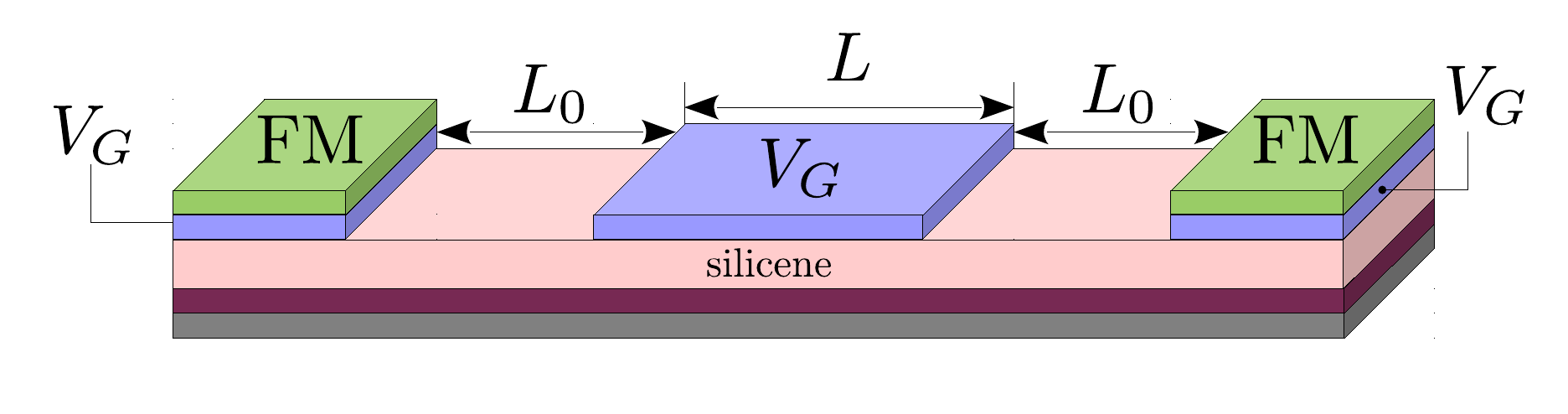}\\
\includegraphics[clip,trim=0cm 0cm -3cm 0cm,width=0.9\textwidth]{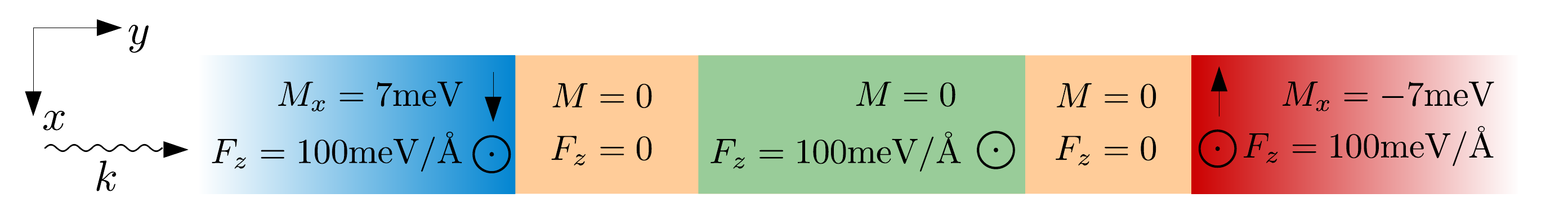}\\
\includegraphics[clip,trim=0cm 1.4cm 2.2cm 0cm,scale=0.5]{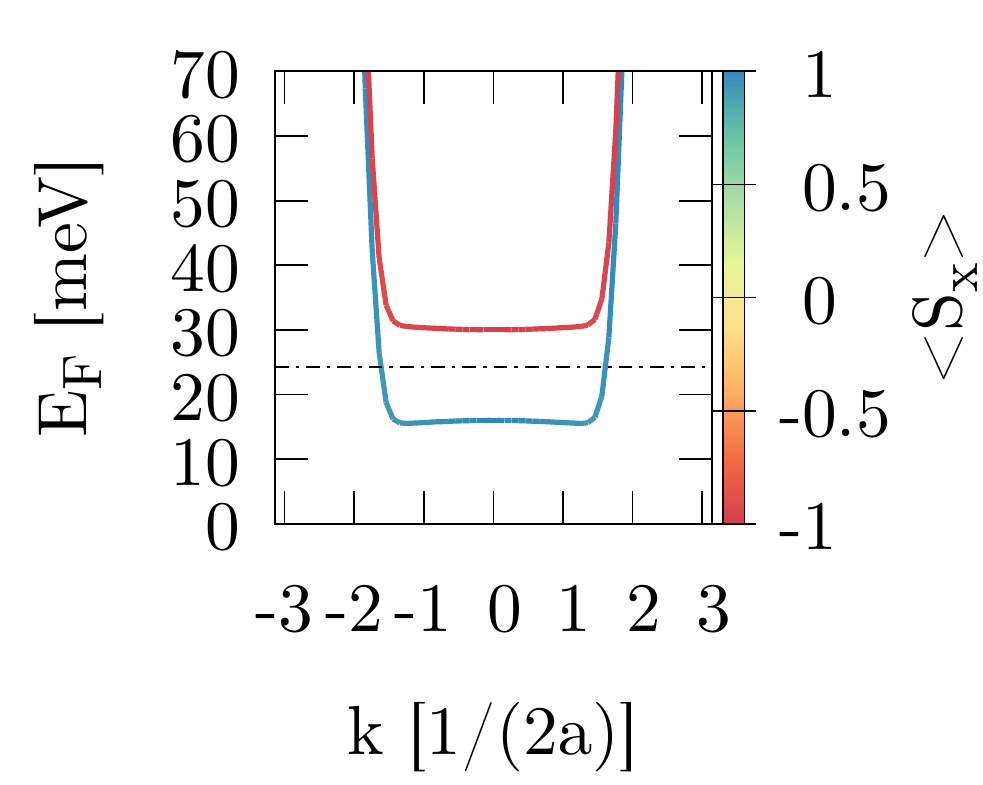}\put(-45,90){(a)}\put(-50,-10){k $\big[\frac{1}{2a}\big]$}
\includegraphics[clip,trim=2.8cm 1.4cm 2.2cm 0cm,scale=0.5]{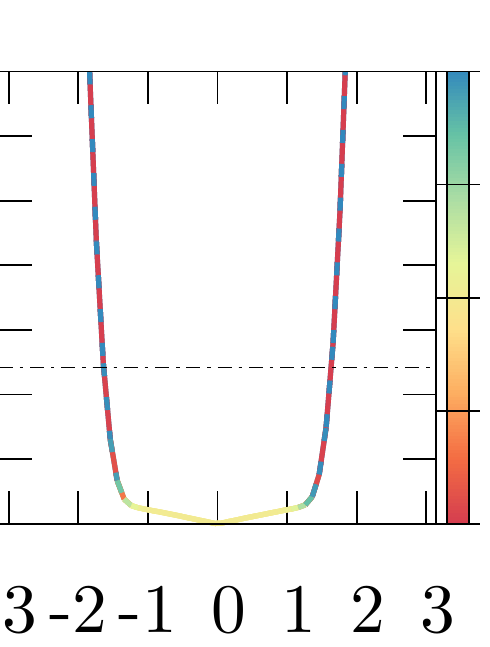}\put(-45,90){(b)}\put(-50,-10){k $\big[\frac{1}{2a}\big]$}
\includegraphics[clip,trim=2.8cm 1.4cm 2.2cm 0cm,scale=0.5]{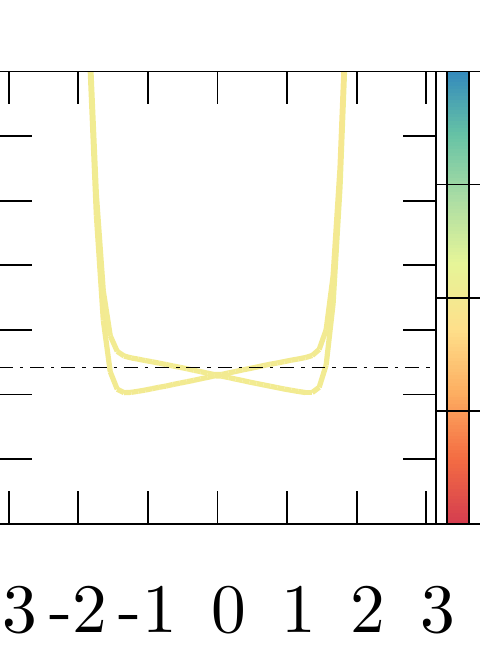}\put(-45,90){(c)}\put(-50,-10){k $\big[\frac{1}{2a}\big]$}
\includegraphics[clip,trim=2.8cm 1.4cm 2.2cm 0cm,scale=0.5]{bandfm_cc.pdf}\put(-45,90){(d)}\put(-50,-10){k $\big[\frac{1}{2a}\big]$}
\includegraphics[clip,trim=2.8cm 1.4cm 0cm 0cm,scale=0.5]{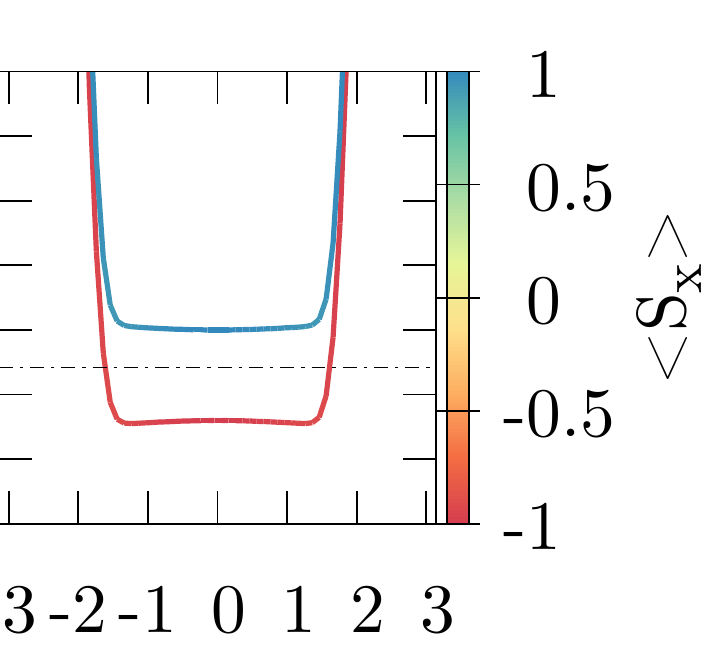}\put(-75,90){(e)}\put(-80,-10){k $\big[\frac{1}{2a}\big]$}
\caption{Schematic view of a spin detector with FM placed above the left (a) and right (e) lead of silicene nanoribbon with gate voltage $F_z = 100$ meV/\AA. Sections (b) and (d) correspond to pristine silicene and (c) is a gated area where precession occurs ($L=5.5$ nm). Below each section (a-e) band structure of its infinite counterpart has been drawn.  Horizontal line denotes for the energy $E_F = 24.2$ meV accordingly to the previous spin-flip length.  $L_0$ is the spacer between the gated region and the 
leads.  }
\label{fig:FM-sch}
\end{figure*}

\begin{figure}[htbp]
\centering
\includegraphics[width=0.4\textwidth]{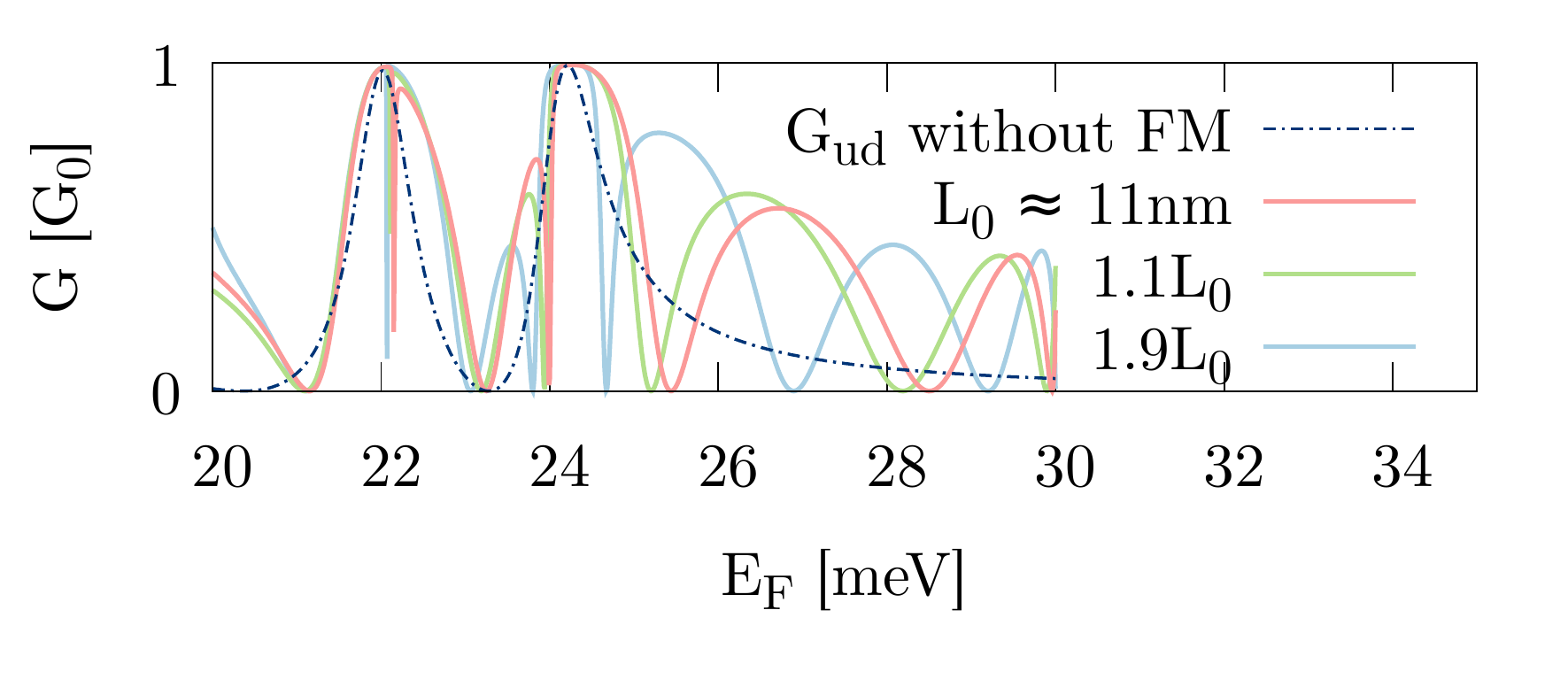}\\

\caption{Total conductance $G$ (solid lines) as function of the spacer length $L_0$ (see Fig. \ref{fig:FM-sch}(b,d)) compared to the $G_{ud}$ (dashed line) in nanoribbon without FM leads (see Fig. \ref{fig:4-trans}). }
\label{fig:FMtrans}
\end{figure}

\begin{figure}[htbp]
\centering
\includegraphics[width=0.33\textwidth]{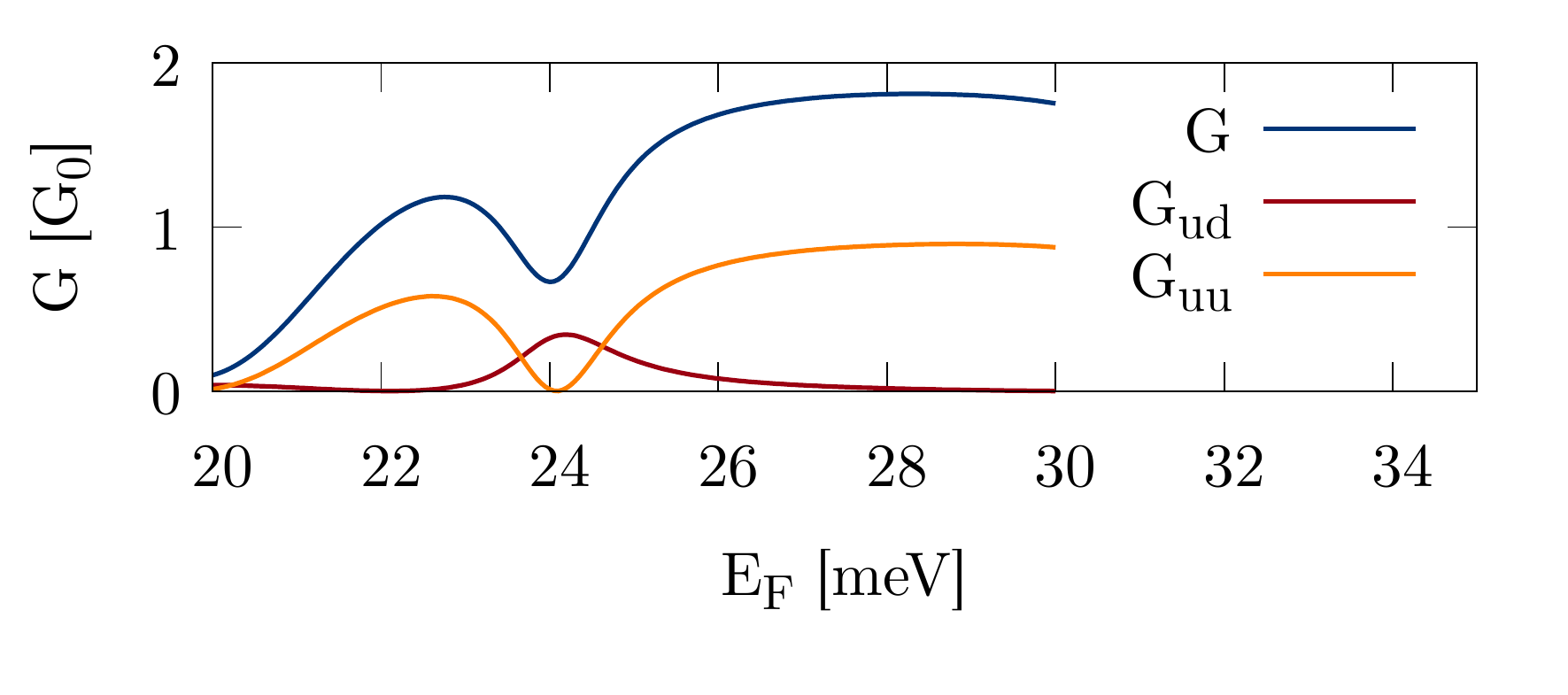}
\includegraphics[width=0.14\textwidth]{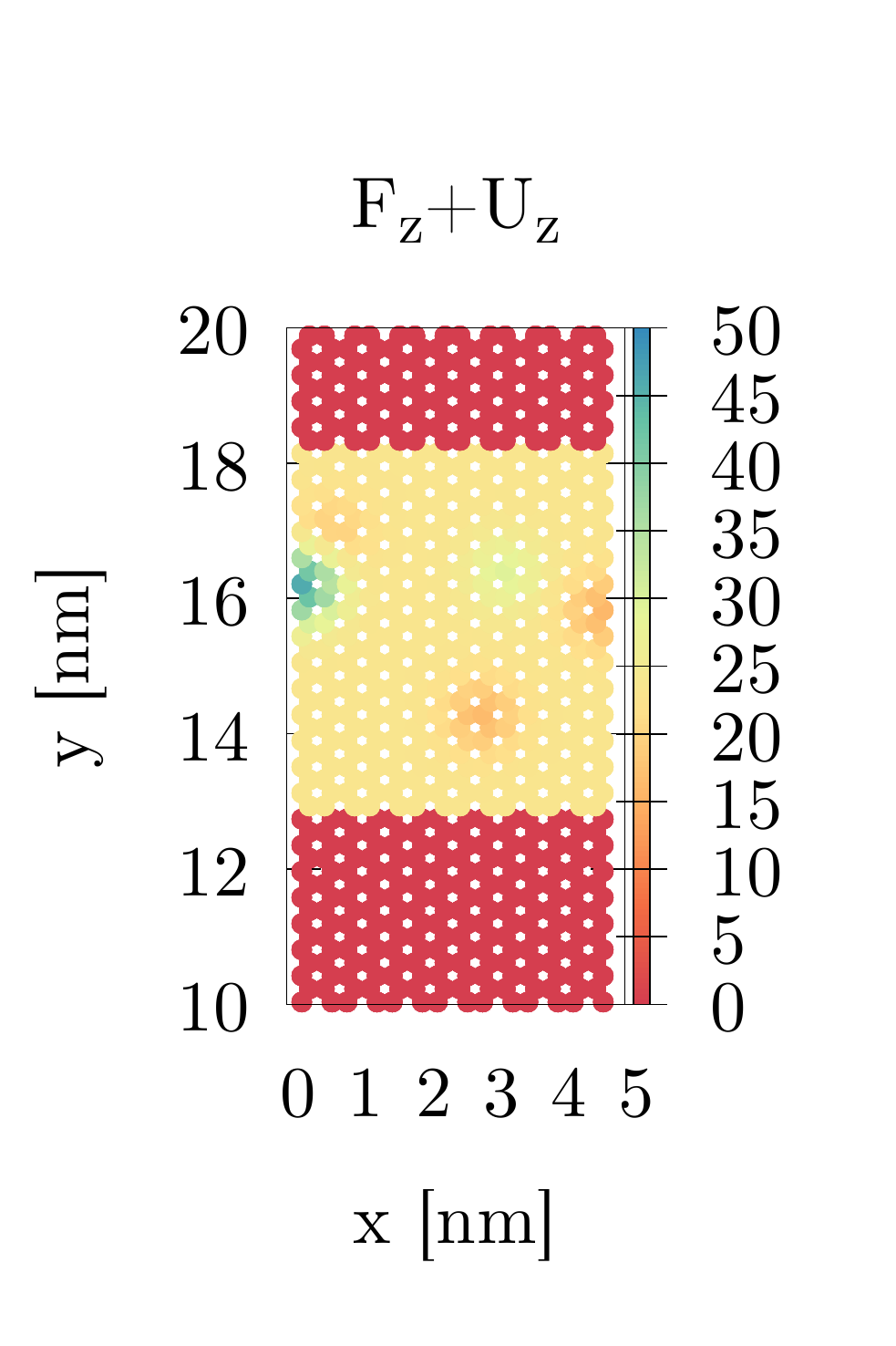}\put(-230,80){(a)}\put(-60,90){(b)}\put(-5,40){\rotatebox{90}{\tiny $|V_z|$ [meV]}}\\
\includegraphics[width=0.33\textwidth]{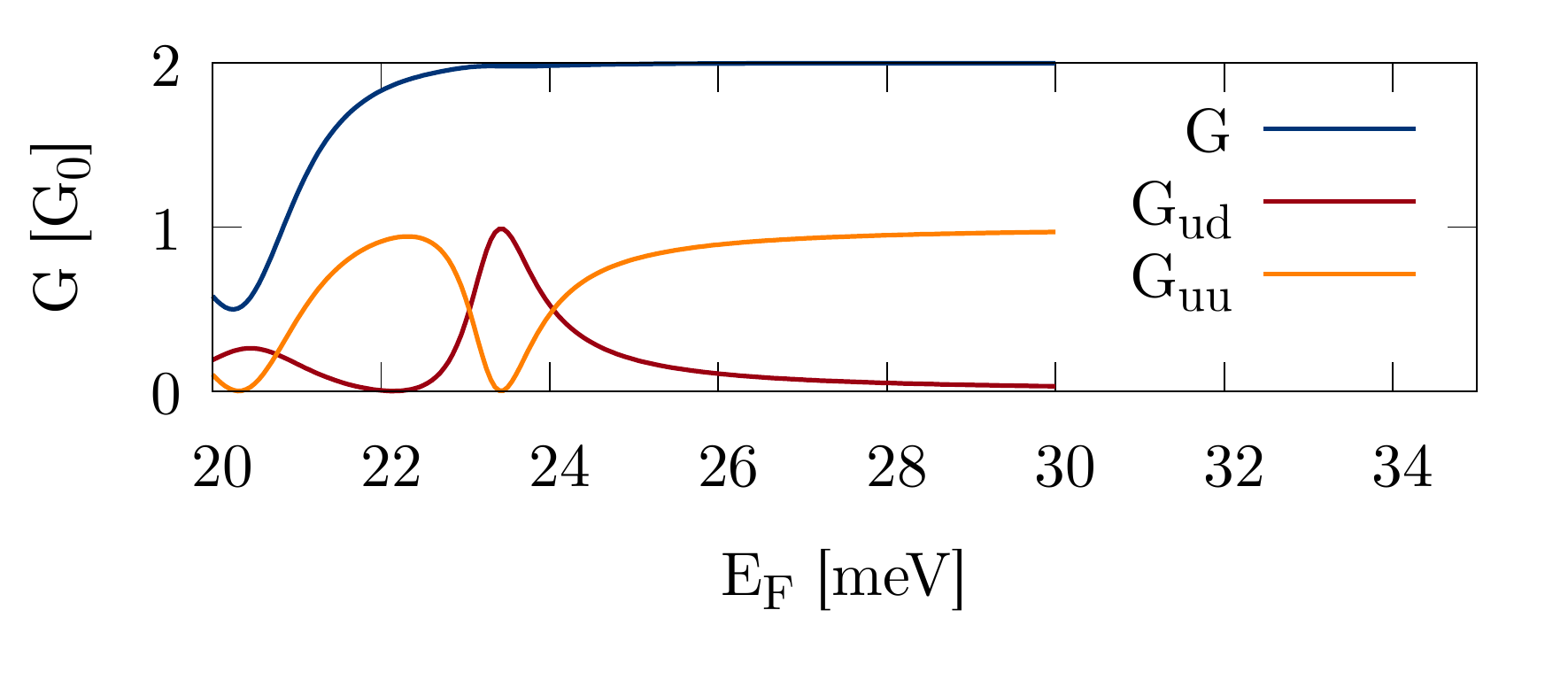}
\includegraphics[width=0.14\textwidth]{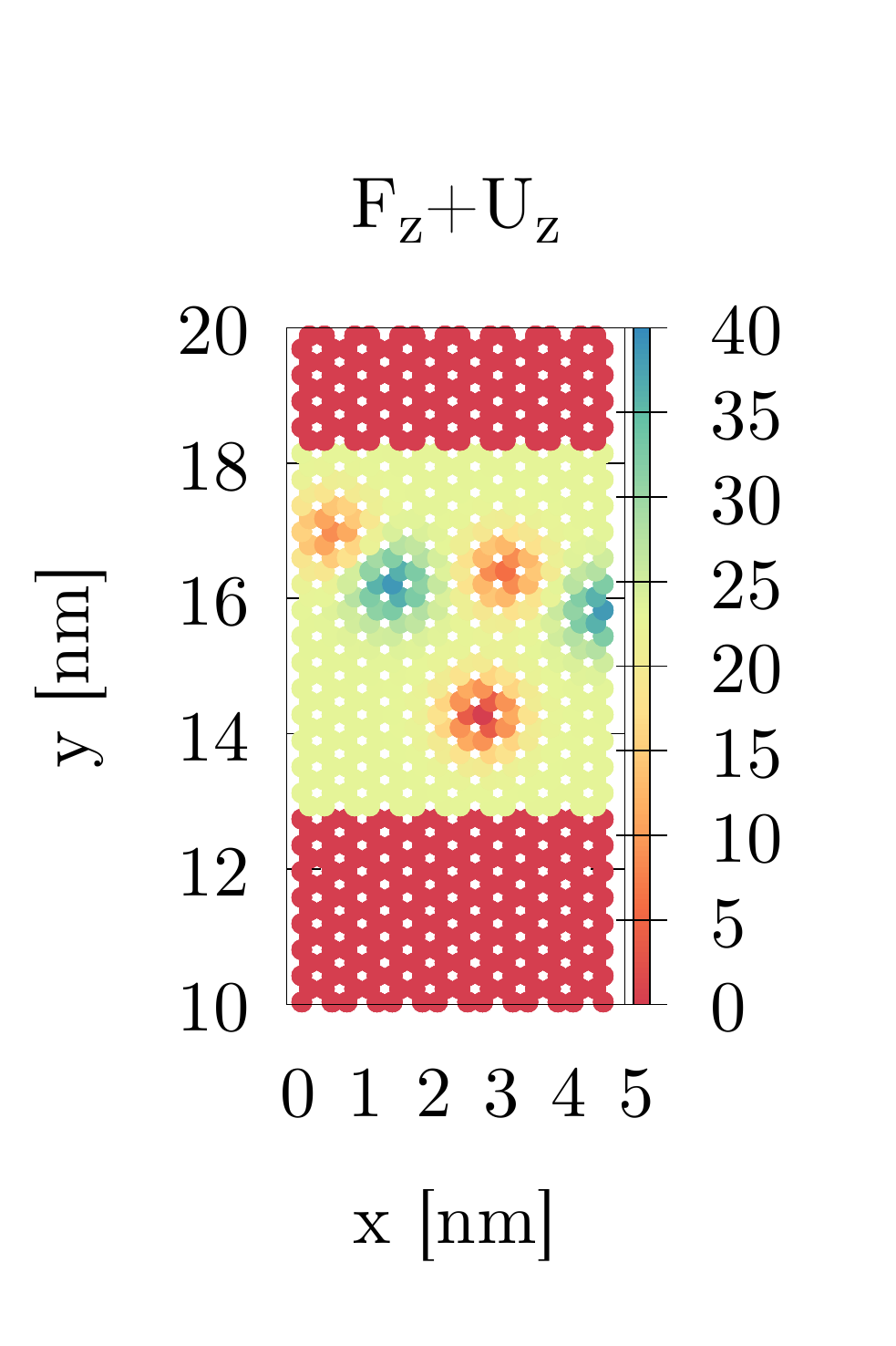}\put(-230,80){(c)}\put(-60,90){(d)}\put(-5,40){\rotatebox{90}{\tiny $|V_z|$ [meV]}}
\caption{ Conductance $G$ with spin conserving $G_{uu}$ and spin-flipping $G_{ud}$ parts in disordered system for a random fluctuation of the vertical field $U_z$ 
including  the left edge (a), and  with no fluctuation on the left edge (c). Map of on-site energies that correspond to $\vert V_z \vert \equiv  e(F_z+U_z) \vert \ell_k \vert$ has been presented for each case (b) and (d), respectively. Magnitude of the fluctuating field $U_z$ in both cases is comparable to $F_z$ and corresponds to energy $\approx$ 20meV. }
\label{fig:fluct}
\end{figure}
%

\subsection{Spin-flip detection}

Above, we indicated the setup for the in-plane spin inversion. Let us now
discuss a possible experimental setup that could detect the inversion. 
For that purpose we consider a system with ferromagnetic insulators placed above the silicene. Due to the proximity effect the exchange energy $\mathbf{M}$ appears in Hamiltonian (\ref{eq:h0})  and modifies the band structure. In Fig. \ref{fig:FM-sch}(a,e) we plotted the dispersion relation for $M_x=\pm 7$ meV and $F_z=100$ mV/\AA{} that can be achieved in a proximity of a ferromagnet magnetized along the $x$ axis [see
the sketch on top of Fig. \ref{fig:FM-sch}]. 

The configuration proposed in Fig. \ref{fig:FM-sch} filters the spins polarized along 
the $x$ axis by choosing the Fermi energy that corresponds to only one conductive subband. Spin polarization along the $+x$ or $-x$ direction for that propagation is allowed on each lead  and can be set by an adequate FM polarization.
The length of the spacer $L_0$ was set to 11 nm and the length of the top gate [Fig. \ref{fig:FM-sch}(c)] in the middle remains the same as in the system without FM, $L=5.5$ nm, for the band structure of Fig. \ref{fig:4-agressive}.

The setup of Fig. \ref{fig:FM-sch}  works in the following way:
the total conductance of the system is equal to 1 flux quantum only if the full spin-flip occurs [Fig. \ref{fig:FMtrans}] and the maxima of $G$ coincide
with the maxima of $G_{ud}$ that were obtained in Fig. 7 without the ferromagnets. Figure \ref{fig:FMtrans} presents the results for varied length of the spacer $L_0$. Although the conductance varies with $L_0$, the total conductance stays at 1 $G_0$
for the Fermi energies that correspond to the spin flip under the gate of
the fixed length $L$. 
The proposed device  should allow to tune the Fermi energy to obtain a perfect spin inverter.

Note, that for the spin injection by the input lead alternatively to the magnetic proximity effect one can use the procedure for all-electrical generation of spin-polarized currents by an energy-dependent phase difference for the electron spin
proposed recently \cite{tao17}.

\subsection{Electrostatic disorder}

In order to check the robustness of a spin-flipping conductance  to a disorder we
returned to the system without the FM proximity effect and we introduced inhomogeneities
to the vertical electric field that produces Gaussian impurities in the potential \cite{fluct}
\begin{equation}
U_z = \sum_{i}^{N}  U_i \exp{(-\vert{} \mathbf{r_k}-\mathbf{R_i} \vert{}^2/2\eta^2)},
\end{equation} 
\noindent where $\mathbf{R_i}$ is the center of the $i$-th impurity and $\eta$ is set equal to the lattice constant. 
 $U_i \in (-\mathfrak{m}F_z,\mathfrak{m}F_z) $ is a scale factor  chosen randomly and multiplied by magnitude  factor $\mathfrak{m}$ equal 1 or 0.1.
This fluctuations is introduced in our calculations by 
modification of the vertical electric field $F^d_z(x,y)=F_z+U_z(x,y))$ in Hamiltonian $H_0$ (\ref{eq:h0}). 

We considered $N=5$ impurities in  the  gated area. When the magnitude of $U_i$ was set to $\mathfrak{m}=0.1$ we observe only a slight difference in the spin-flipping conductance. The results is  still very similar to the case without the fluctuations [Fig. \ref{fig:4-trans}]. For magnitude $\mathfrak{m}=1$ we can distinguish two cases: (i) when one of the impurity center is localized at the left edge of the nanoribbon then spin-flip conductance drastically falls down [Fig. \ref{fig:fluct}(a,b)] because most of wave function goes along left edge and any fluctuation on this path just blocks the electron propagation
(ii) when all impurities centers are localized inside top gate excluding left edge the one-inversion conductance is still available but spin-flipping conductance with more than one inversion along path is hampered [Fig. \ref{fig:fluct}(c,d)].

\section{Summary and conclusions}
We considered a gated segment of a silicene nanoribbon as a spin inverter via precession of the incident spin in the effective magnetic field due to the SO interactions.
The Rashba interaction due to the external electric field fails to induce the spin inversion in the zigzag ribbon for which the intrinsic SO interaction
keeps the incident electron spin polarized along the $z$ direction. The perpendicular polarization of the electron spin is not present for the
armchair ribbon that allows the Rashba interaction to drive the spin-precession. However, the resulting spin precession length is large, of the order of $\mu$m. 
We demonstrated that the gated zigzag nanoribbon can be used as an inverter of in-plane polarized incident spins and that spin precession length can be very short
for low Fermi energy, e.g.  less than 10 nm for a reasonable value of the external electric field $F_z=100$ meV/\AA. That spin inversion length strongly depends of the $\Delta k$ chosen through the $E_F$ level that can be tuned in a large range due to the spin splitting in the band structure when the external electric field is applied.
With the  local exchange field it is possible to prepare spin-flip detection device that is transparent for the spin-polarized transport only for a perfect spin inversion
in the gates area.
Electric field fluctuations can suppress the precession if the center of impurity appears on the left edge of the nanoribbon and its magnitude is comparable to applied external electric field. In other cases precession is hampered but spin-flip is still observed.

\section*{Acknowledgments}
We would like to thank Alina Mre\'n{}ca-Kolasi\'n{}ska for helpful discussions.
B.R. is supported by Polish government budget for science in 2017-2021 as a research project under the program  "Diamentowy Grant" (Grant No. 0045/DIA/2017/46)
and by the EU Project POWER.03.02.00-00-I004/16. B.S. acknowledges the support of NCN grant DEC-2016/23/B/ST3/00821. The calculations were performed on PL-Grid Infrastructure.

\bibliographystyle{apsrev4-1}
\bibliography{bib_silicene}

\end{document}